\documentclass[aps,twocolumn,superscriptaddress]{revtex4-1}

\usepackage[pdftex]{graphicx}
\usepackage{subfigure}
\usepackage{hyperref}
\usepackage{units}
\usepackage{amsmath}
\usepackage{color}

\begin{document}

\title{Fermi surface properties of the bifunctional organic metal  $\kappa$-(BETS)$_2$Mn[N(CN)$_2$]$_3$ near
the metal--insulator transition}

\author{V.N. Zverev}
\affiliation{Institute of Solid State Physics, Russian Academy of Sciences,
Academician Ossipyan Str. 2, Chernogolovka, 142432 Russia}
\affiliation{Moscow Institute of Physics and Technology, Institutskii
9, Dolgoprudny, 141700, Russia}


\author{W. Biberacher}
\affiliation{Walther-Mei{\ss}ner-Institut, Bayerische Akademie der
Wissenschaften, Walther-Mei{\ss}ner-Strasse 8, D-85748 Garching, Germany}

\author{S. Oberbauer}
\affiliation{Walther-Mei{\ss}ner-Institut, Bayerische Akademie der
Wissenschaften, Walther-Mei{\ss}ner-Strasse 8, D-85748 Garching, Germany}
\affiliation{Physik-Department, Technische Universit\"{a}t M\"{u}nchen,  James-Franck-Stra{\ss}e 1, 85748 Garching, Germany}

\author{I. Sheikin}
\affiliation{Laboratoire National des Champs Magn\'{e}tiques Intenses, (LNCMI-EMFL), CNRS, UJF, 38042 Grenoble, France}

\author{P. Alemany}
\affiliation{Departament de Ci\`{e}ncia de Materials i Qu\'{i}mica F\'{i}sica and Institut de Qu\'{i}mica Te\'{o}rica i Computacional (IQTCUB), Universitat de Barcelona, Mart\'{i} i Franqu\`{e}s 1, 08028 Barcelona, Spain}

\author{E. Canadell}
\affiliation{Institut de Ci\`{e}ncia de Materials de Barcelona (ICMAB-CSIC), Campus de la UAB, 08193 Bellaterra, Spain}

\author{M.V. Kartsovnik}
\email[]{mark.kartsovnik@wmi.badw.de}
\affiliation{Walther-Mei{\ss}ner-Institut, Bayerische Akademie der
Wissenschaften, Walther-Mei{\ss}ner-Strasse 8, D-85748 Garching, Germany}

\begin{abstract}
We present detailed studies of the high-field magnetoresistance of the layered organic metal $\kappa$-(BETS)$_2$\-Mn\-[N(CN)$_2$]$_3$ under a pressure slightly above the insulator-metal transition. The experimental data are analysed in terms of the Fermi surface properties and compared with the results of first-principles band structure calculations. The calculated size and shape of the inplane Fermi surface are in very good agreement with those derived from Shubnikov-de Haas oscillations as well as the classical angle-dependent magnetoresistance oscillations. A comparison of the experimentally obtained effective cyclotron masses with the calculated band masses reveals electron correlations significantly dependent on the electron momentum. The momentum- or band-dependent mobility is also reflected in the behavior of the classical magnetoresistance anisotropy in a magnetic field parallel to layers. Other characteristics of the conducting system related to interlayer charge transfer and scattering mechanisms are discussed based on the experimental data. Besides the known high-field effects associated with the Fermi surface geometry, new pronounced features have been found in the angle-dependent magnetoresistance, which might be caused by coupling of the metallic charge transport to a magnetic instability in proximity to the metal-insulator phase boundary.
\end{abstract}

\date{\today}
\maketitle

\section{Introduction}
\label{intro}

The organic charge transfer salt $\kappa$-(BETS)$_2$\-Mn\-[N(CN)$_2$]$_3$, where BETS stands for
bis\-(ethylene\-dithio)\-tetra\-selena\-fulvalene,
belongs to the family of hybrid molecular
conductors which can be seen as natural multilayer structures of conducting and magnetic layers
alternating on a subnanometer scale \cite{koba04,coro04,ouah12}. While itinerant $\pi$ electrons in the
BETS donor layers are responsible for metallic conduction, the magnetic moment is mainly determined
by localized $d$-electron spins of Mn$^{2+}$ ions in the insulating anion layers \cite{kush08}.
On the other hand, the quasi-two-dimensional (quasi-2D) conducting system undergoes a metal-insulator
transition at $T_{\mathrm{MI}}\simeq 23$\,K presumably associated with the Mott instability \cite{kush08,zver10}.
The insulating ground state is suppressed by a quasi-hydrostatic
pressure of about 1\,kbar, giving way to a metallic and even superconducting state with
$T_c \approx 5.5$\,K. However, the shape of the ``pressure--temperature'' phase diagram of
this compound \cite{zver10} significantly differs from that of archetypical organic Mott
insulators $\kappa$-(BEDT-TTF)$_2$X with anions X$^-$ = \{Cu[N(CN)$_2$]Cl\}$^-$ and
\{Cu$_2$(CN)$_3$\}$^-$ \cite{lefe00,kaga04a,kuro05}.
A reason for that may lie in the interaction between the conduction
$\pi$-electrons and localized $d$-electron spins. The $\pi$-$d$ exchange coupling is known to be at the core
of the metal-insulator transition of another hybrid organic salt $\lambda$-(BETS)$_2$FeCl$_4$
\cite{bros98,cepa02}. Moreover, in some BETS salts with Fe-cointaining tetrahedral anions
the $\pi$-$d$ coupling is clearly manifest already in the metallic state, playing, for example, a crucial
role in stabilizing superconductivity in a magnetic field \cite{koba04,uji01a,uji03b,fuji02,kono04b,kart16}.
In the present material this coupling seems to be considerably weaker. It has been found to
cause changes of magnetic properties of the Mn$^{2+}$ subsystem upon entering the insulating
state \cite{vyas11,vyas12,vyas17}. However, no evidence of its influence on the conducting system has
been reported so far.

For a better understanding of the mechanisms of the insulating and superconducting instabilities
a thorough knowledge of the Fermi surface properties is indispensable.
To that end, we have carried out a detailed study of the high-field magnetoresistance
of pressurized $\kappa$-(BETS)$_2$Mn[N(CN)$_2$]$_3$ supplemented by first-principles band structure
calculations.
For most of the measurements the pressure value
of $p \approx 1.4$\,kbar was chosen so as to drive the compound into the fully normal state, but not far away
from the metal-insulator phase boundary \cite{zver10}.
In fields above 12\,T quantum (Shubnikov-de Haas, SdH) oscillations have been found, providing a direct
access to the topology and size of the 2D Fermi surface. Further quantitative information on the size and
shape of the Fermi surface has been obtained from the classical angle-dependent magnetoresistance oscillations
(AMRO).
Besides the detailed Fermi surface geometry, the SdH data and classical magnetoresistance yield other important
characteristics of the conducting system such as effective cyclotron masses of the charge carriers,
scattering parameters, and interlayer transfer energy.
In particular, by confronting the effective mass values obtained from the experiment with the calculated band masses
we find a quite strong, momentum-dependent renormalization effect caused by electron correlations in the proximity
to the metal-insulator transition. Finally, in addition to the ``conventional'' phenomena determined by the geometry
of the quasi-2D Fermi surface, the classical magnetoresistance has shown new features, which may be related to
an interaction of the charge carriers with the magnetic subsystem.

The paper is organized as follows. In the next Section the experimental details are described.
Section \ref{band} presents the results of the first-principles band structure calculations. The predicted
conducting bands and Fermi surface look similar to those obtained by the semi-empirical extended H\"{u}ckel
method. However, the density of states and relevant band cyclotron masses are 40\% higher. This rather large
difference, often found in the organics, see, e.g., \cite{meri00a}, is important to take into account
when estimating the strength of many-body renormalization effects on the experimentally determined mass.
Section \ref{Sect_MQO} presents experimental data on the SdH oscillations and their analysis.
In Sect.\,\ref{Sect_MR} the behavior of the classical component of the interlayer magnetoresistance as
a function of the strength and orientation of magnetic field is considered. A summary and concluding remarks
are given in Sect.\,\ref{summary}.

\section{Experimental}
\label{exper}
The samples used in the experiments were electrochemically grown single crystals \cite{kush08} with
typical dimensions $\sim 0.5 \times 0.3 \times 0.02$\,mm$^3$, the largest dimensions being in the
plane of conducting layers, that is, the crystallographic $bc$-plane.
Electrical leads for four-probe resistance measurements were made by attaching annealed 20\,$\mu$m thick Pt
wires to the sample surface using a conducting graphite paste. All the measurements were done in the
interlayer resistance geometry, which is the most convenient and informative for layered organic
conductors, see, e.g., \cite{kart04} for a review. The resistance was measured by the standard
low-frequency a.c. technique. Field-dependent magnetoresistance with the focus on the SdH oscillations
was measured in the temperature interval 0.36 to 1.0 K at a current of 1\,$\mu$A assuring no overheating
for sample resistance values $< 3$\,k$\Omega$.
Angle-dependent measurements were done in liquid $^4$He at $1.3-1.4$\,K  with a current of 10\,$\mu$A.

Quasihydrostatic pressure was applied using a BeCu clamp cell with silicon oil
as a pressure medium. A calibrated manganin coil with a resistance of $\approx 6$\,$\Omega$ and
sensitivity 0.243\%/kbar was used as a resistive pressure gauge. All the measurements, except one
run presented at the end of Sect.\,\ref{Sect_mass}, were done at pressure $ p \approx 1.4$\,kbar.

All the field sweeps and most of the angle-dependent magnetoresistance data presented in the paper
were carried out in a 30\,T resistive magnet at the LNCMI-Grenoble. 15\,T angular sweeps shown in
Sect.\,\ref{AMR} and some test measurements were done using a superconducting solenoid.

For the angle-dependent studies the samples were mounted on a two-axes rotating stage. Continuous
rotations in different planes perpendicular to the plane of conducting layers were done at a fixed
field strength. The sample orientation was defined by polar angle $\theta$ between the field direction
and the normal to the layers and by azimuthal angle $\varphi$ between the field projection on the layer
plane and the crystallographic (inplane) $c$-axis. The angular resolution was $< 0.1^{\circ}$ and
$\approx 0.5^{\circ}$ for $\theta$ and $\varphi$, respectively. The initial orientation of the sample
was set with an error bar of $\simeq \pm 3^{\circ}$ for both $\theta$ and $\varphi$. However, by using
the center of the dip in the $R(\theta)$ dependence as a reference point for the exact inplane field direction
($|\theta| = 90^{\circ}$), the $\theta$ error bar was reduced to $< 0.5^{\circ}$.

Three high-quality samples were used in the experiments, all showing consistent data both on
quantum oscillations and on the classical magnetoresistance. In what follows, we will present detailed
data obtained on two different samples, respectively, from field sweeps in the orientation
perpendicular to the layers and from the angular sweeps at a fixed field strength.

\section{First-principles calculation of the conduction bands}
\label{band}

\begin{figure}[tb]
    \centering
        \includegraphics[width=0.45\textwidth]{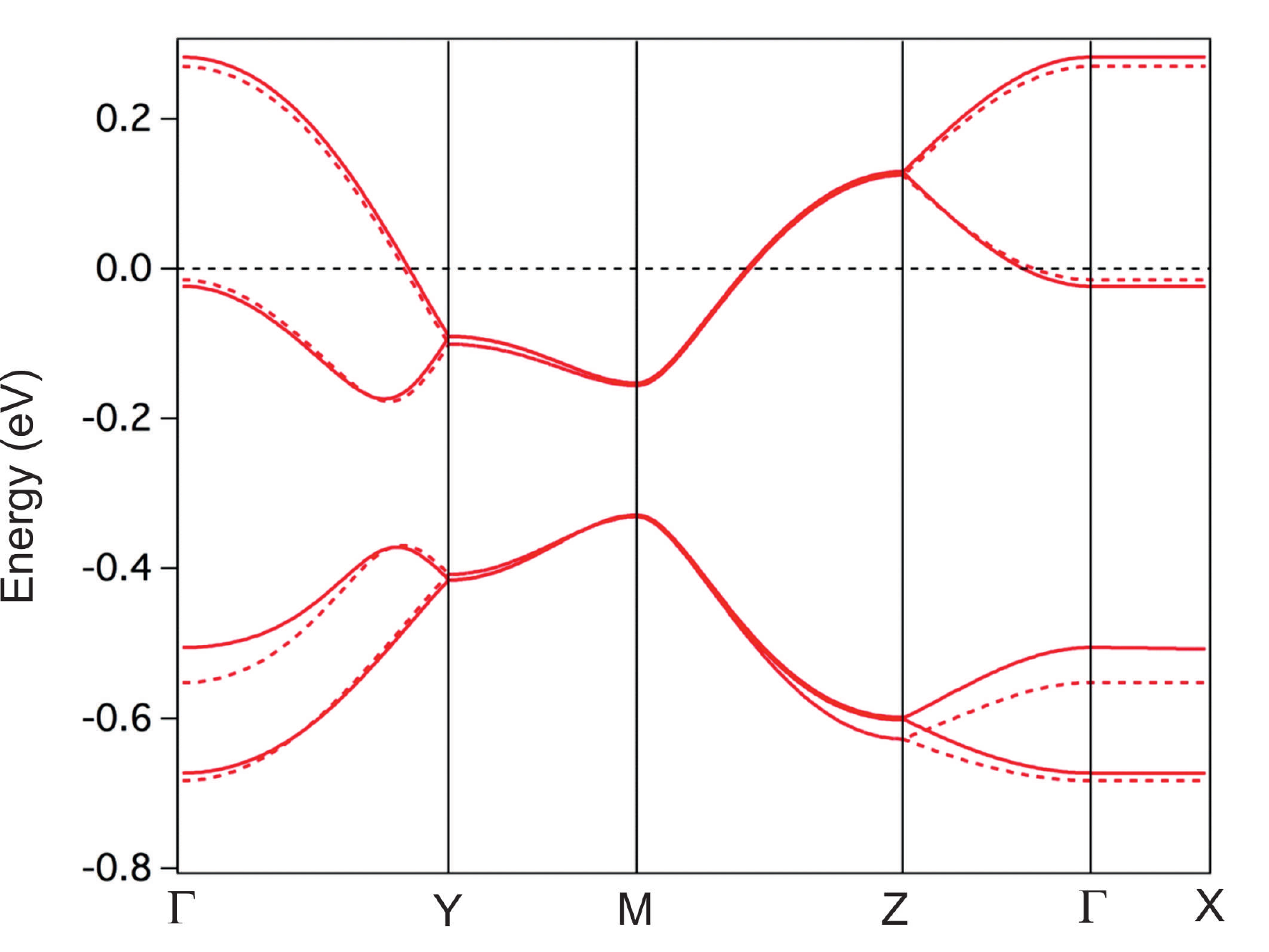}
    \caption{(Color online) Calculated band structure of $\kappa$-(BETS)$_2$Mn[N(CN)$_2$]$_3$ based on the 15\,K crystal structure \cite{zver10}. The result of the full calculation is shown with solid lines, whereas the result of the calculation where the anions were replaced by a uniform background of charge is shown with dashed lines. The energy is counted from the Fermi level. $\Gamma = (0, 0, 0)$, $\mathrm{X} = (1/2, 0, 0)$, $\mathrm{Z} = (0, 0, 1/2)$ and $\mathrm{M} = (1/2, 0, 1/2)$ in units of the monoclinic reciprocal lattice vectors. }
    \label{Bands}
\end{figure}

Calculations of the low-temperature band structure were carried out
using a spin-polarized numerical atomic orbitals density functional
theory (DFT) approach \cite{hohe64,*kohn65} in the generalized
gradient approximation (GGA) \cite{perd96}. Only the valence
electrons were considered in the calculations with the core being
replaced by norm-conserving scalar relativistic pseudopotentials
\cite{trou91} factorized in the Kleinman-Bylander form
\cite{klei82}. We have used a split-valence double-$\zeta$ basis set
including polarization orbitals with an energy shift of 10 meV for
all atoms \cite{arta99}. The energy cutoff of the real space
integration mesh was 350\,Ry. The Brillouin zone was sampled using a
grid of $(5\times 20\times 20)$ $k$-points \cite{monk76} in the
irreducible part of the Brillouin zone. The experimental crystal
structure at 15\,K \cite{zver10} was used in the calculations. The
calculated bands near the Fermi level are shown in
Fig.\,\ref{Bands}.
They contain only contributions from the highest occupied molecular orbitals of BETS and have shapes typical of strongly dimerized $\kappa$-salts of BEDT-TTF and BETS.
The width of the bands crossing the Fermi level is 0.46\,eV. This value is lower than that obtained by the extended H\"{u}ckel method, 0.65\,meV \cite{zver10}, as it is often found for organic charge transfer salts. As discussed below, the weaker dispersion leads to higher values of the density of states and cyclotron masses. On the other hand, comparing to the values $0.40\pm 0.02$\,eV obtained by first-principles calculations \cite{kand09,xu95,chin97,jesc12} for $\kappa$-(BEDT-TTF)$_2$X salts, exhibiting the Mott-insulating instability, the present value is very similar, just slightly higher.
Along the interlayer direction ($\Gamma$\,--\,\-X) the dispersion is below the resolution of our calculations.

\begin{figure}[tb]
    \centering
        \includegraphics[width=0.35\textwidth]{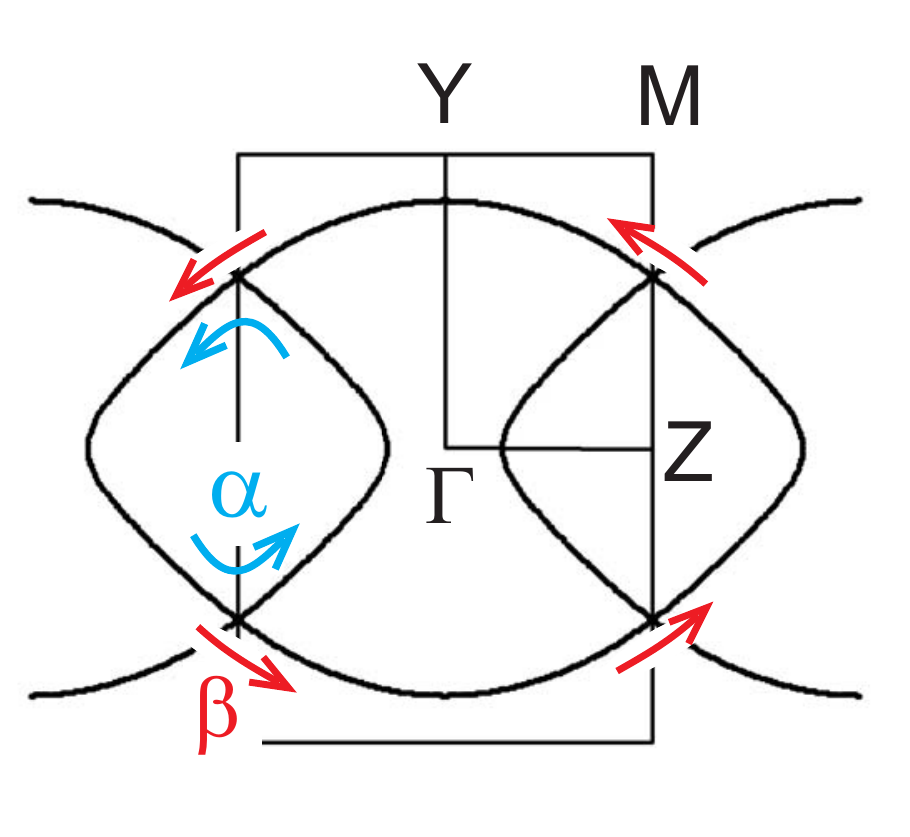}
    \caption{(Color online) 2D Fermi surface of $\kappa$-(BETS)$_2$Mn[N(CN)$_2$]$_3$. Thin lines indicate the principal directions of the reciprocal lattice and the first Brillouin zone boundary. The arrows show the directions of the cyclotron motion on the classical (blue) and magnetic-breakdown (red) orbits in a magnetic field.}
    \label{FS}
\end{figure}

The calculated 2D Fermi surface is shown in Fig.\,\ref{FS}. It is a cylinder crossing the Brillouin zone boundary along Z\,--\,\-M. As expected, it shares all features of the Fermi surface obtained by the extended  H\"{u}ckel method \cite{zver10}, in particular, the presence of a rhombus-like portion around point $Z$ with quite flat (however, slightly more rounded near $\Gamma$) sides. The area of this rhombus-like part is $25.2 \%$ of the Brillouin zone cross section.

Due to the crystal symmetry, the calculated two upper bands in Fig.\,\ref{Bands} are degenerate along Z\,--\,\-M, which causes crossing of the adjacent Fermi surfaces on the Brillouin zone boundary.
It should be noted, however, that our DFT calculations do not take into account a statistical disorder of the dicyanamide groups of the anion along the crystallographic $b$-axis \cite{kush08}. This disorder barely affects the electronic structure of the donor layer, however the associated random potential lifts the double degeneracy of the crystal orbitals along the Z\,--\,\-M boundary of the Brillouin zone. As a result, small gaps arise between the rhombus-like Fermi pocket and the open sheets extended along the Y\,--\,\-M direction. As will be shown in the next Section, the presence of the gaps is confirmed by magnetic quantum oscillations. The oscillations reveal a classical cyclotron orbit on the rhombus-like pocket (conventionally labeled as $\alpha$-orbit) and a large magnetic-breakdown orbit ($\beta$-orbit) caused by tunneling through the gaps and encircling the entire Fermi surface, as illustrated in Fig.\,\ref{FS}.

\begin{figure}[tb]
    \centering
        \includegraphics[width=0.45\textwidth]{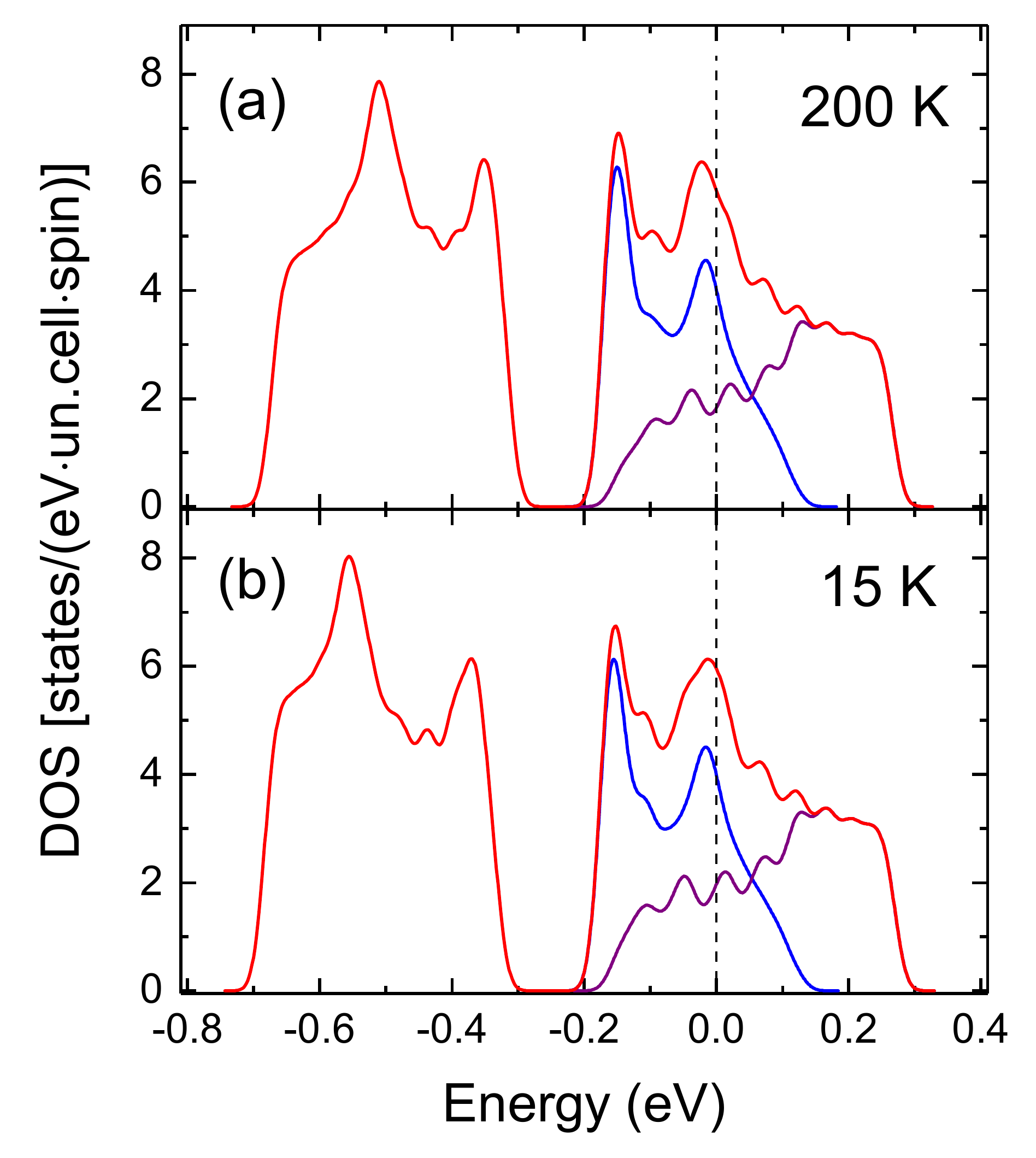}
    \caption{(Color online) Calculated DOS per spin per unit cell for $\kappa$-(BETS)$_2$Mn[N(CN)$_2$]$_3$ at (a) $T = 200$\,K and (b) $T = 15$\,K. The energy is counted from the Fermi level. The contribution from the lower lying band, associated with the rhombus-like Fermi pocket around point Z in Fig.\,\ref{FS}, is shown in blue. The contribution from the higher lying band is shown in purple. The total DOS is represented by the red line.}
    \label{DOS}
\end{figure}

Shown in Fig.\,\ref{DOS} is the density of states (DOS) calculated for two temperatures.
The blue curve is the contribution of the lower-lying partially filled band, which forms the rhombus-like portion of the FS, and the purple curve corresponds to the upper band associated with the open sheets  extended along the Y--M direction. The total DOS is given in red. Interestingly, by contrast to other $\kappa$-type salts, the upper part of the DOS exhibits two pronounced peaks and the Fermi level occurs very near the top of one of them. Qualitatively the same result was obtained by the extended H\"{u}ckel method \cite{zver10} and attributed to a significant inplane anisotropy of the present salt: the coupling between chains of dimers, running along the crystallographic $b$ axis, is weaker than the intrachain interactions. This anisotropy causes a flattening of the lower  partially filled band around point $\Gamma$, near the Fermi level. The resulting peak in the DOS shifts even more close to the Fermi energy at decreasing temperature, as one can see from Fig.\,\ref{DOS}.

Knowing the DOS, one can evaluate the cyclotron mass $m_c$ on the Fermi surface. For a quasi-2D metal there is a simple relation between the two quantities \cite{meri00a}:
\begin{equation}
m_{c} = 2\pi\hbar^2\mathcal{D}_0/(bc) ,
\label{mc}
\end{equation}
where $\mathcal{D}_0$ is the 2D DOS (per spin per unit cell) at the Fermi level, and $b = 8.35$\,{\AA} and $c=11.83$\,{\AA} are the unit cell parameters in the plane of conducting layers \cite{zver10}.
Substituting in Eq.\,(\ref{mc}) the calculated values $\mathcal{D}_{0,\beta} = 5.94$\,eV$^{-1}$ and $\mathcal{D}_{0,\alpha} = 3.98$\,eV$^{-1}$ for the total DOS and for the contribution from the rhombus-like Fermi pocket, respectively, we obtain the cyclotron masses $m_{c,\beta} = 2.89 m_e$ and $m_{c,\alpha} = 1.93 m_e$, where $m_e$ is the free electron mass. These values are 1.4 times larger than the masses ($2. 03m_e$ and $1.39m_e$, respectively) following from the extended H\"{u}ckel calculations \cite{zver10}. However, the ratio $m_{c,\alpha}/m_{c,\beta} = 0.67$ is almost the same. Note that this ratio is $30\%$ higher than what one usually obtains, both theoretically and experimentally, for $\kappa$-salts \cite{meri00a,kawa06a}. The reason for this obviously lies in the fact that the enhancement of the DOS at the Fermi level originates solely from the band responsible for the $\alpha$ pocket.

\section{Magnetic quantum oscillations}
\label{Sect_MQO}
\begin{figure}
    \centering
        \includegraphics[width=0.45\textwidth]{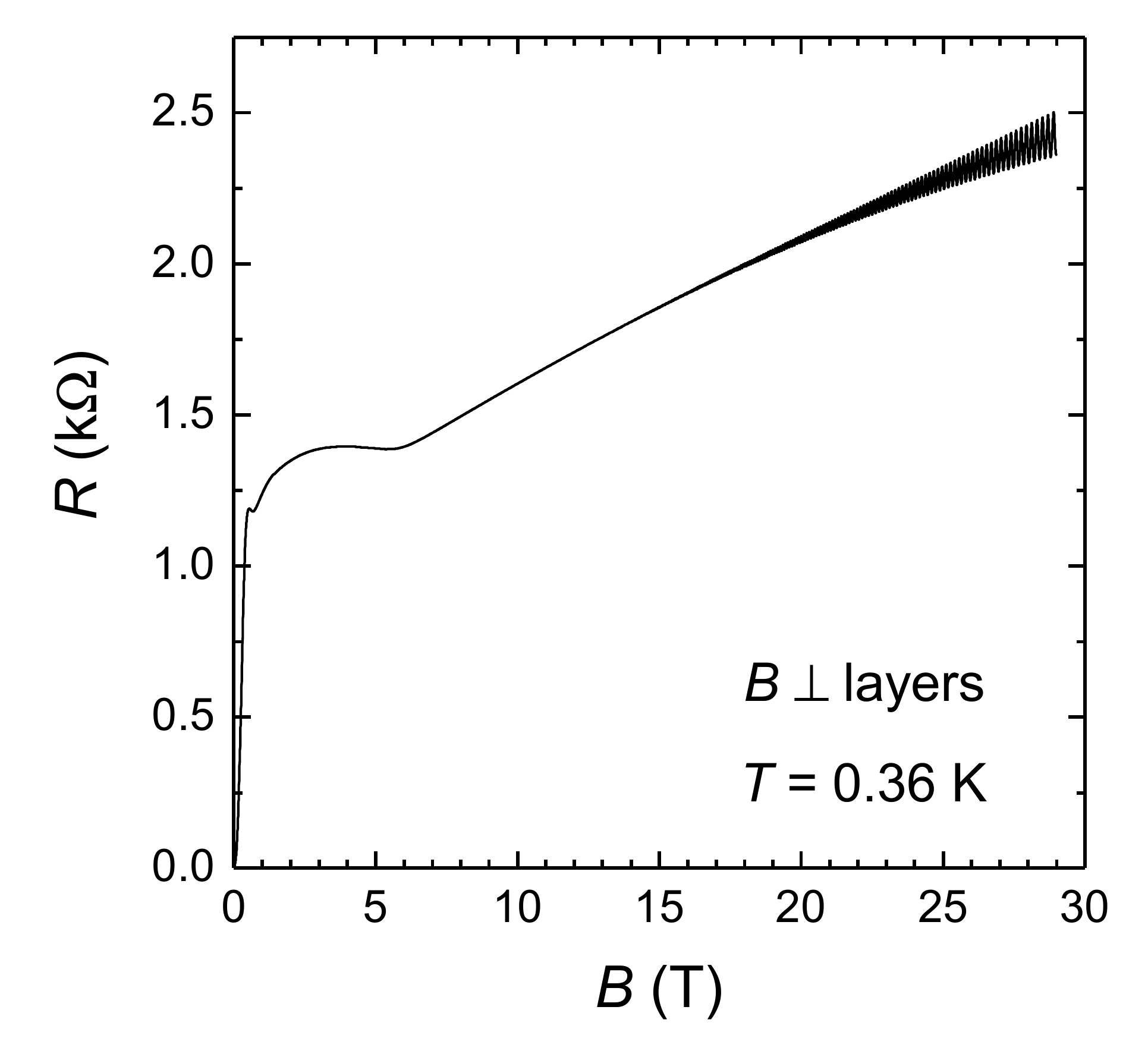}
    \caption{Interlayer resistance of $\kappa$-(BETS)$_2$Mn[N(CN)$_2$]$_3$ under
    a pressure of 1.4\,kbar, at $T = 0.36$\,K,
    as a function of magnetic field normal to layers.}
    \label{R(B)}
\end{figure}
Figure\,\ref{R(B)} shows the general behavior of the interlayer
resistance of pressurized $\kappa$-(BETS)$_2$\-Mn\-[N(CN)$_2$]$_3$
at a temperature $T = 0.36$\,K, in a magnetic field perpendicular to layers.
Besides superconductivity at very low fields, the magnetoresistance
exhibits a few features which will be addressed in the following.
We start with a detailed consideration of the Shubnikov-de Haas (SdH)
oscillations observed at fields $B \gtrsim 12$\,T.

\subsection{SdH spectrum and the Fermi surface topology}
\label{Sect_SdH}
\begin{figure}
    \centering
        \includegraphics[width=0.45\textwidth]{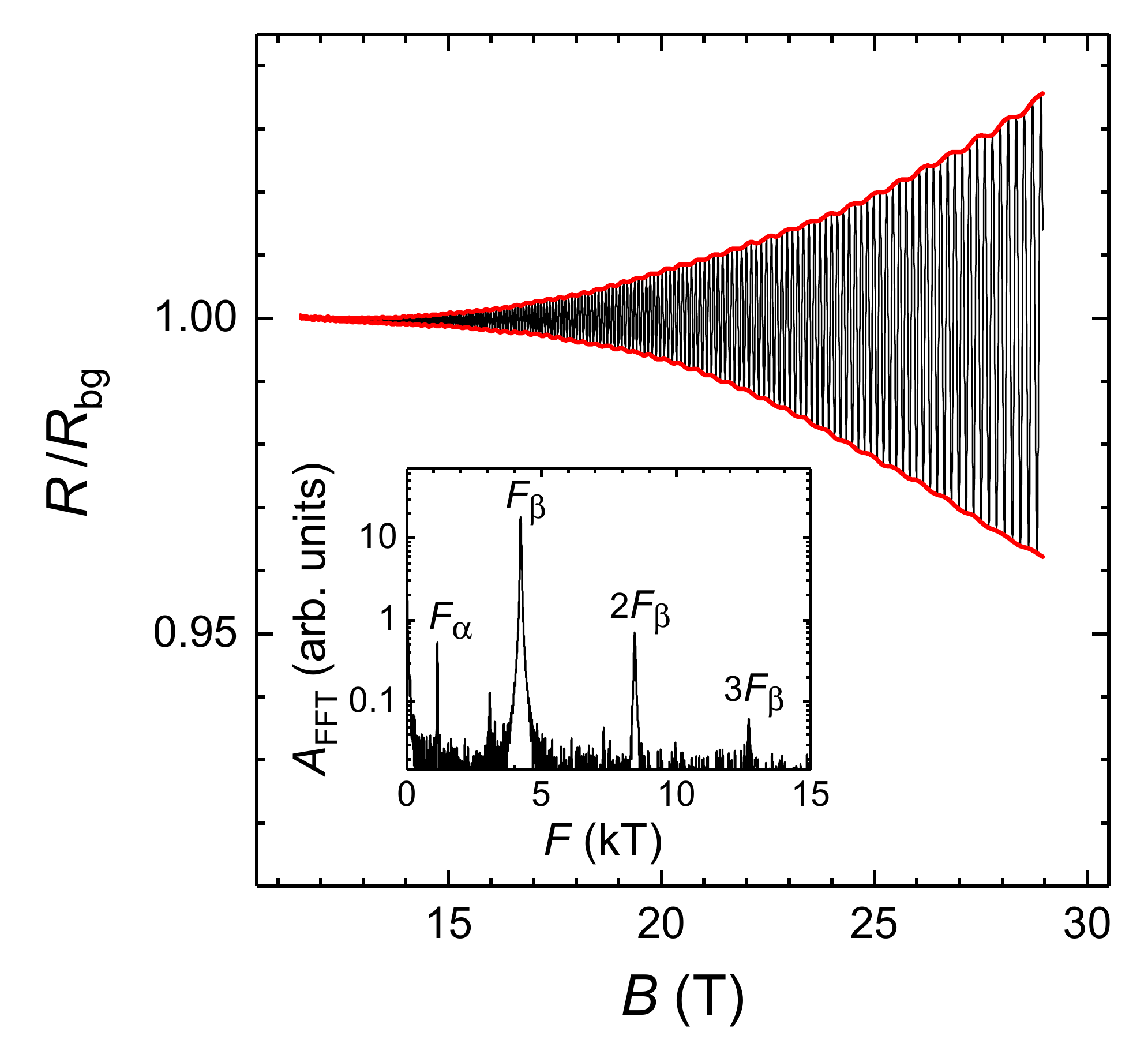}
    \caption{(Color online) Oscillatory component of the field-dependent resistance in Fig.\,\ref{R(B)} normalized to the nonoscillating background. Red lines are envelopes of the rapid $\beta$-oscillations originating from magnetic breakdown, to emphasize weak slow oscillations associated with the classical cyclotron orbit $\alpha$, see Fig.\,\ref{FS}. Inset: the corresponding FFT spectrum.}
    \label{SdH}
\end{figure}
An example of the oscillatory part of magnetoresistance
is presented in Fig.\,\ref{SdH}.
It is dominated by rapid SdH oscillations; the fast Fourier
transform (FFT) spectrum, shown in the inset, has a peak at a
frequency $F_{\beta}=4225$\,T. The relevant cyclotron orbit $\beta$
in $\mathbf{k}$-space covers the area $S_{\beta} = 40.31$\,nm$^{-2}$. This
coincides, within an accuracy of $1\%$, with the first Brillouin zone area
calculated from the 15\,K crystallographic data \cite{zver10}.
In addition, weak slower oscillations can be resolved in the envelopes of the main oscillations (red lines in Fig.\,\ref{SdH}).
The corresponding peak in the FFT spectrum is at $F_{\alpha} =
1135$\,T, revealing  a cyclotron orbit area of $10.83$\,nm$^{-2}$ or
$27.1\%$ of the Brillouin zone area. This agrees fairly well with the size of the
rhombus-like part of the calculated 2D Fermi surface centered at
point $Z$ on the Brillouin zone boundary, see Fig.\,\ref{FS}. The
presence of this oscillatory component indicates that there is no
band degeneracy at the zone boundary: the Fermi surface consists of
a pair of open sheets and a cylinder separated from each other by a
small gap. The slow oscillations originate from the classical orbit
$\alpha$ on the Fermi cylinder indicated by the blue arrows in
Fig.\,\ref{FS}, whereas the fast oscillations are a result of
magnetic breakdown (MB) through the gaps (red arrows in
Fig.\,\ref{FS}).

While not predicted by band structure calculations, a small MB gap
between the open sheets and cylindrical Fermi surface has also been
found in SdH experiments on several other $\kappa$-type salts of
BETS and BEDT-TTF with a center-symmetric layer structure
\cite{bali00,*uji01c,kono05,harr98b,togo01,helm95,*harr98,kart95c,yama96,weis99b}.
One can consider a weak, $\lesssim 1$\,meV, spin-orbit interaction
as a possible source of the gap \cite{wint17}. However, as pointed
out in Sec.\,\ref{band}, in the present case a gap should already
arise due to the disorder in the anion layer along the $b$-axis.

In the earlier study \cite{zver10} performed at similar pressures no
SdH oscillations with the frequencies $F_{\alpha}$ and $F_{\beta}$
were observed, but instead a very low frequency $F_{\gamma} = 88$\,T
has been found and attributed to a very small Fermi pocket. The
latter could be formed due to folding the original Fermi surface
caused by the superstructure transition at about 100\,K
\cite{zver10}. In that scenario the orbits $\alpha$ and $\beta$ can
also be realized, but now both would additionally require magnetic
breakdown through the superstructure gap. The absence of the
relevant frequencies could be attributed to the lower field range,
$B\leq 15$\,T, and higher temperatures, $T\geq 1.4$\,K, used in the
experiment \cite{zver10}. On the other hand, the reason for the
absence of $F_{\gamma}$ in our present data is not quite clear. It
is possible that the discrepancy is caused by different pressurizing
procedures applied in the two experiments. In the work \cite{zver10}
the sample was cooled at ambient pressure down to low temperatures
and pressures of $\sim 1$\,kbar were applied below 20\,K using the
helium gas pressure technique. In the present experiment the sample
was first pressurized at room temperature in the clamp cell and then
cooled down under pressure. One can speculate that the 100\,K
transition responsible for the Fermi surface reconstruction is
suppressed under these conditions, which would explain the absence
of the slow oscillations in our data. To verify this scenario, it
would be interesting to perform low-temperature X-ray studies at
different pressures.

\subsection{Effective cyclotron masses}
\label{Sect_mass}
The effective cyclotron masses corresponding to the $\alpha$ and $\beta$ orbits can be evaluated in the standard way from the temperature dependence of the oscillation amplitude.
 The latter is described by the Lifshitz-Kosevich temperature damping factor $R_{T,\alpha(\beta)}$ \cite{lifs55,shoe84}:
\begin{equation}
R_{T,\alpha(\beta)} = \frac{K\mu_{\alpha(\beta)} T/B}{\sinh(K\mu_{\alpha(\beta)} T/B)}\,,
\label{RT}
\end{equation}
where
\begin{equation}
K=2\pi^2 k_{\mathrm{B}} m_e / \hbar e \approx 14.69\,\mathrm{T/K},
\label{K}
\end{equation}
$k_{\mathrm{B}}$ is the Boltzmann constant, $e$ the elementary charge, and $\mu_{\alpha(\beta)} = m_{c, \alpha(\beta)}/m_e$, the cyclotron mass on the $\alpha$($\beta$) orbit expressed in free electron mass units.
\begin{figure}[tb]
    \centering
        \includegraphics[width=0.45\textwidth]{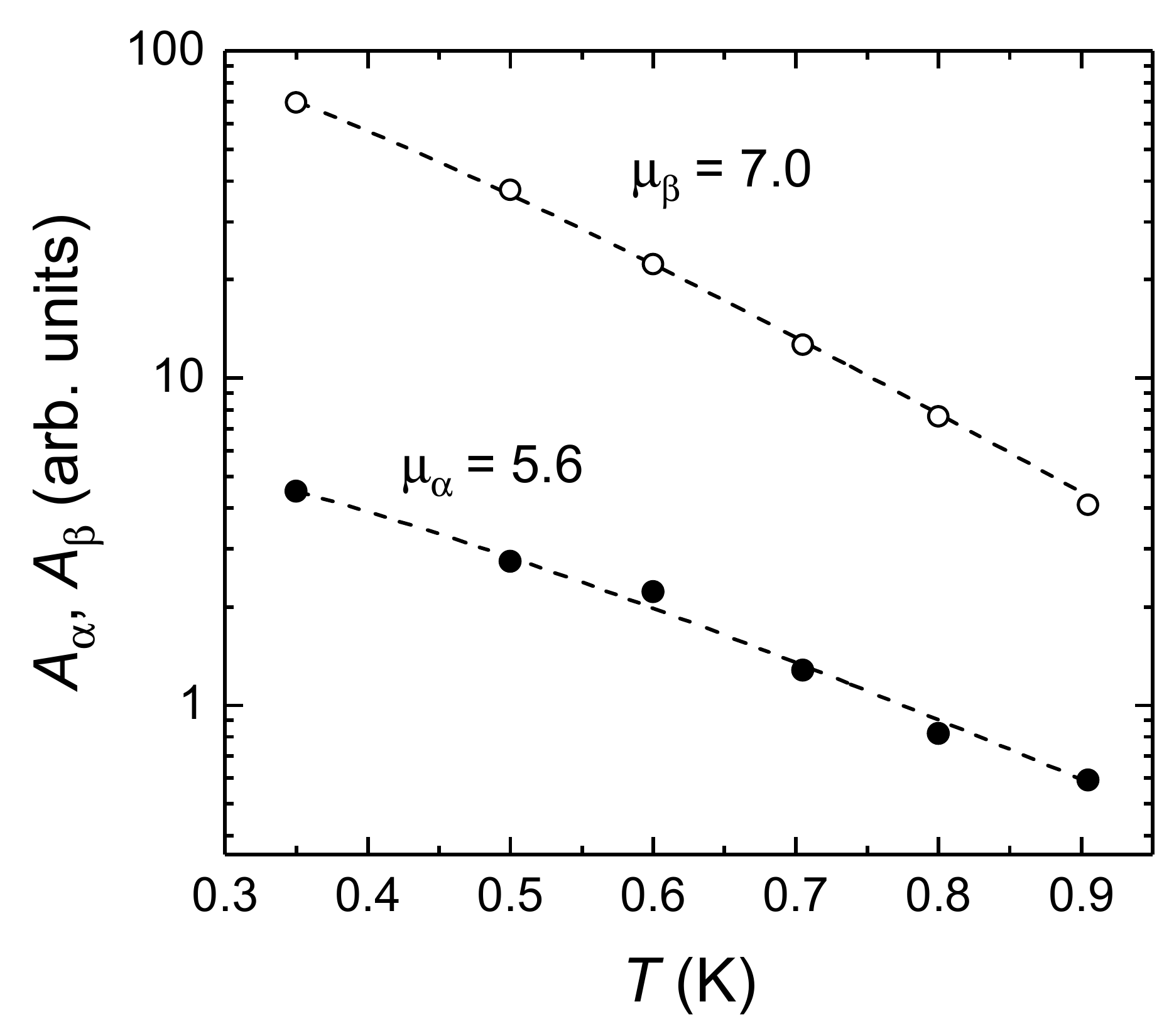}
    \caption{Temperature dependence of the FFT amplitudes of the $\alpha$ (filled symbols) and $\beta$ (open symbols) oscillations.
    The lines are fits to the Lifshitz-Kosevich temperature dependence given by Eq.\,(\ref{RT}) with the normalized cyclotron mass
    $\mu_{\alpha(\beta)} = m_{c, \alpha(\beta)}/m_e$ as a fitting parameter.}
    \label{mass}
\end{figure}
Figure\,\ref{mass} shows the FFT amplitudes of the $\alpha$ and $\beta$ oscillations obtained in the field window from 14 to 17\,T at different temperatures. Fitting the experimental data by the Lifshitz-Kosevich temperature dependence yields the effective cyclotron masses $\mu_{\alpha} = 5.6\pm0.1$ and $\mu_{\beta} = 7.0\pm0.05$.

The experimentally determined cyclotron masses significantly exceed the theoretical values given in Sect.\,\ref{band}. This apparent discrepancy is often observed for the $\kappa$-type salts and attributed to many-body effects \cite{meri00a,kawa06a}: electron-electron and electron-phonon interactions lead to a renormalization of the effective mass $\mu$ entering Eq.\,(\ref{RT}) by a factor
$r>1$ as compared to the ``band'' mass $\mu^{\mathrm{b}}$ obtained from the band structure calculations (here we defined $\mu^{\mathrm{b}}$ as the mass obtained from Eq.\,(\ref{mc}) and normalized to the free electron mass) \cite{shoe84}.
However, usually the renormalization is uniform over the Fermi surface, i.e. the factor $r$ is the same for the $\alpha$ and $\beta$ orbits \cite{meri00a,kawa06a}. By contrast, in our case the renormalization factor for the $\alpha$ orbit, $r_{\alpha} \equiv \mu_{\alpha}/\mu^{\mathrm{b}}_{\alpha} = 2.9$, is notably higher than for the $\beta$ orbit, $r_{\beta} \equiv\mu_{\beta}/\mu^{\mathrm{b}}_{\beta} = 2.4$. Keeping in mind that the $\beta$ orbit contains {\em all} the states on the Fermi surface, including those on the $\alpha$-pocket (see Fig.\,\ref{FS}), the difference between the many-body renormalization on the $\alpha$ orbit and on the rest of the Fermi surface must be even stronger.

The reason for the enhanced many-body effects on the $\alpha$ pocket may be qualitatively understood by taking into account the proximity of the electronic system to the metal-insulator transition. As shown in Sect.\,\ref{band}, the band associated with the $\alpha$ pocket is partially flattened around the Fermi level (which already causes an increase of the one-particle band mass $\mu^{\mathrm{b}}_{\alpha}$). It is reasonable to expect that the effective reduction of the bandwidth places this part of the conduction system more close to the Mott-insulating state, resulting in a relative enhancement of electron correlation effects. Additionally, the rhombus-like shape of the $\alpha$ pocket is suggestive of the so-called ``nesting'' instability, that is, a strongly enhanced scattering at the wave vector connecting the opposite flat segments of the pocket \cite{whan91,bori08,ruva91}. This may further contribute to the many-body renormalization factor for the effective mass.

To check the role of the proximity to the Mott-insulating state, we have repeated the SdH experiment at an elevated pressure, $p = 4.1$\,kbar, moving the material far away from the metal-insulator phase boundary. Expectedly, the cyclotron masses become considerably smaller at this pressure: $\mu_{\alpha}(4.1\,\mathrm{kbar}) = 3.4\pm0.1$ and $\mu_{\beta}(4.1\,\mathrm{kbar}) = 5.2\pm0.1$, indicating weakening of the many-body effects. But an important result is that, within the experimental accuracy \cite{comment_pressure}, the mass enhancement factor is now the same for both orbits, $r_{\alpha} = r_{\beta} = 1.78\pm0.05$. This provides a strong support for the suggested above momentum- or band-dependent enhancement of electronic correlations near the metal-insulator transition in $\kappa$-(BETS)$_2$Mn[N(CN)$_2$]$_3$. It is worth noting that even at this pressure the mass values are relatively high compared to the other $\kappa$-salts \cite{caul94,weis99a,togo01,bigg02}. This can be at least partially attributed to the peak in the one-particle DOS near the Fermi level predicted by the band structure calculations, see Sect.\,\ref{band}.

\subsection{Field dependence of the SdH amplitudes}
\label{Sect_B}
As one can see in Fig.\,\ref{SdH}, the oscillation amplitude increases in a monotonic manner with no traces of beating. The absence of beats suggests that the Landau level separation near the Fermi energy, $\hbar\omega_c$
(where $\omega_c = eB/m_c$ is the cyclotron frequency), is larger than the interlayer bandwidth, $4t_{\perp}$, in the whole field range where the oscillations are observed. Therefore we apply the 2D
Lifshitz-Kosevich-Shoenberg formula \cite{shoe84,shoe84a,cham02b,grig11},
\begin{equation}
A_i(B) = A_{0,i}R_{T,i} R_{\mathrm{D},i} R_{\mathrm{MB},i}, \qquad  i = \alpha, \beta ,
\label{Ai}
\end{equation}
which is valid for weak oscillations in a 2D system for analysing the magnetic field dependence of the oscillation amplitudes. Besides the temperature factor introduced above, this formula contains the Dingle damping factor $R_{\mathrm{D}}$, determined by scattering, and the MB factor. The prefactor $A_{0,i}$ is proportional to the contribution from the carriers on the $i$-th orbit to the zero-field conductivity, $A_{0,i} \propto \sigma_{0,i}$. It does not depend on $B$, but we have included it in Eq.\,(\ref{Ai}) since it has to be taken into account when comparing the amplitudes of the $\alpha$ and $\beta$ oscillations. The Dingle factor is conventionally considered in the form \cite{shoe84,ding52,bych61}:
\begin{equation}
R_{\mathrm{D}} = \exp\left(-\frac{\pi}{\omega_c \tau}\right) = \exp\left(-K \mu T_{\mathrm{D}}/B\right),
\label{RD}
\end{equation}
where $K$ is defined by Eq.\,(\ref{K}), and the Dingle temperature $T_{\mathrm{D}} = \hbar/2\pi k_{\mathrm{B}}\tau$ is associated with the scattering rate $1/\tau$.
The MB factors for orbits $\alpha$ and $\beta$ are readily expressed in the form (see, e.g., \cite{sasa91a}):
\begin{subequations}
\label{RMB}
\begin{align}
R_{\mathrm{MB},\alpha} & = \left[1-\exp\left(-B_{\mathrm{MB}}/B \right) \right], \label{RMBa} \\
R_{\mathrm{MB},\beta} & =  \exp\left(-2B_{\mathrm{MB}}/B\right) \label{RMBb}
\end{align}
\end{subequations}
with the characteristic field related to the energy gap $\Delta_{\mathrm{MB}}$ at the MB junction \cite{shoe84}: $B_{\mathrm{MB}} \sim (\Delta_{\mathrm{MB}}^2/\varepsilon_{\mathrm{F}})(m_c/\hbar e)$.

\begin{figure}[tb]
    \centering
        \includegraphics[width=0.45\textwidth]{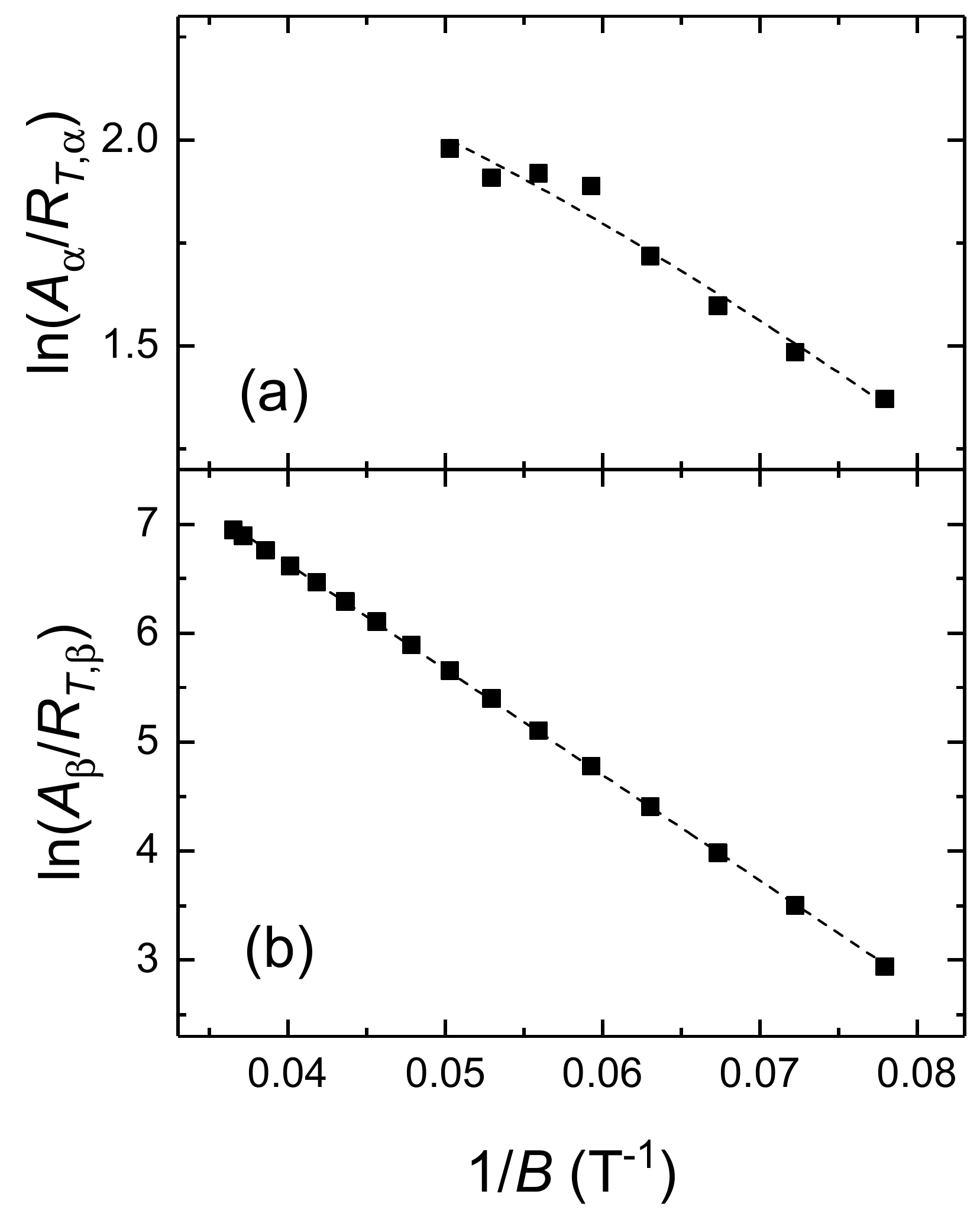}
    \caption{Dingle plots for the amplitudes of the $\alpha$ oscillations (a) and the $\beta$ oscillations (b), see text.
The dashed lines are fits based on Eqs.\,(\ref{Ai})-(\ref{RMB}).
}
    \label{Dingle}
\end{figure}

Figure \ref{Dingle} shows the Dingle plot of the oscillation amplitudes: the experimentally obtained amplitudes, divided by the temperature damping factors (which are known from the above analysis of the $T$-dependence), are plotted in logarithmic scale against inverse magnetic field.
The amplitudes were taken from FFT spectra made in 3\,T-wide field
windows. The horizontal positions of the points correspond to the
midpoints of the respective windows in the $1/B$ scale.

Before starting with the fitting procedure, it should be noted that a precise evaluation of the MB field from our experiment can hardly be done. On the one hand, the functional $B$-dependence of the MB factor for the $\beta$ oscillations is the same as the $B$-dependence of the Dingle factor, cf. Eqs.\,(\ref{RD}) and (\ref{RMBb}). Therefore, one cannot extract separately the values $B_{\mathrm{MB}}$ and $T_{\mathrm{D}}$ from the $\beta$ oscillations only.
On the other hand, the influence of MB on the shape of the $A_{\alpha}(B)$ dependence is very weak. Indeed, despite the higher cyclotron mass, the $\beta$ oscillations strongly dominate in the whole field range in Fig.\,\ref{SdH}, implying that the MB field is well below this range. Hence, the expression in Eq.\,(\ref{RMBa}) for the MB factor for the $\alpha$ oscillations can be approximated as $R_{\mathrm{MB},\alpha} \approx B_{\mathrm{MB}}/B$. This dependence is much weaker than the exponential dependence of the Dingle damping factor. Therefore, the influence of MB on the $\alpha$ oscillations is basically reduced to that on the absolute amplitude.
Further, the prefactors $A_{0,i}$ can hardly be directly evaluated, as they depend on numerous details of interlayer charge transfer and scattering. However, taking into account that the $\beta$ orbit comprises roughly twice as many states as the $\alpha$ orbit, one can tentatively assume $A_{0,\alpha}/A_{0,\beta} \sim 1/2$.

In spite of the mentioned issues affecting the accuracy of
$B_{\mathrm{MB}}$, our analysis yields some interesting qualitative results.

We begin with fitting the amplitude of the $\beta$ oscillations. In the Dingle plot coordinates we obtain a linear fit [dashed line in Fig.\,\ref{Dingle}(b)] with the $y$-intercept $\ln(A_{0,\beta})=10.51\pm0.02$, in the units of the graph, and the slope contributed by both the Dingle and MB factors, $G_{\beta} = - (K\mu_{\beta} T_{\mathrm{D},\beta} + 2B_{\mathrm{MB}}) = - 96.9 \pm 0.3$\,T.

 Next, we turn to the $\alpha$ oscillations. The slope of the fitting curve in Fig.\,\ref{Dingle}(a) is mainly determined by the Dingle factor, yielding $T_{\mathrm{D},\alpha} = 0.48 \pm 0.02$\,K. As mentioned above, the effect of MB on the shape of the field dependence is very weak. It leads to a barely visible nonlinearity of the Dinlge plot in Fig.\,\ref{Dingle}(a). Obviously, this nonlinearity cannot be unambiguously evaluated within the present experimental accuracy.

Nevertheless, the MB field can be estimated from the absolute value
of the $\alpha$-oscillation amplitude provided the coefficient $A_{0,\alpha}$ is
known. Setting, as suggested above $A_{0,\alpha} = A_{0,\beta}/2$ and
using for $A_{0,\beta}$ the value found by fitting the $\beta$ amplitude,
we obtain a very low MB field: $B_{\mathrm{MB}} = 0.057$\,T. For instance, the
corresponding energy gap, $\Delta_{\mathrm{MB}} \sim 0.3$\,meV, is
more than an order of magnitude lower than in
$\kappa$-(BEDT-TTF)$_2$Cu(NCS)$_2$ \cite{sasa91a,meye95,godd04}.
But the large difference between the two cases is not surprising.
Indeed, the lack of the inversion symmetry of the crystal structure in
the latter compound is expected to produce a much larger gap than the
subtle mechanisms discussed above in relation to our material,
see Sects.\,\ref{band}, \ref{Sect_SdH}. However, it should be kept in mind
that the present estimation of $B_{\mathrm{MB}}$ crucially depends
on the assumed ratio $A_{0,\alpha}/A_{0,\beta}$. A decrease of
this ratio would lead to a proportional increase of the estimated MB
field. Thus, the obtained value of $B_{\mathrm{MB}}$ should only be
considered as a very rough estimate.

Coming back to the $\beta$ oscillations, we see that the term associated with MB provides a negligibly small contribution to the slope of the $B$-dependence. Even a 10-fold increase of the given above $B_{\mathrm{MB}}$ value would only lead to a change of $\sim 1\%$ in $G_{\beta}$. So the evaluation of the Dingle temperature is robust against the uncertainty in the MB field. Using the cyclotron mass $\mu_{\beta} = 7.0$, we find $T_{\mathrm{D},\beta} \approx G_{\beta}/K\mu_{\beta}=0.94$\,K.

The obtained values of $T_{\mathrm{D},\alpha}$ and $T_{\mathrm{D},\beta}$ differ from each other by a factor of $\approx 2$. This apparently comes at odds with momentum-independent scattering commonly assumed for our materials at low temperatures.
The difference can be somewhat reduced by a more accurate consideration of many-body renormalization effects. Strictly speaking, both the cyclotron mass and the Dingle temperature in the expression for $R_{\mathrm{D}}$ in Eq.\,(\ref{RD}) are renormalized. It was shown both theoretically \cite{fowl65,enge70,elli80} and experimentally \cite{pali72} that the effects of electron-phonon interactions on $\mu$ and $T_{\mathrm{D}}$ compensate each other  in a broad field and temperature range. The influence of electron-electron interactions is less studied in this respect. However, it was argued \cite{mart03} that at least for a 2D Fermi liquid the same compensation should be valid for any inelastic process, including electron-electron scattering, as long as the oscillations are weak, i.e. $R_T,R_{\mathrm{D}} \ll 1$. These conditions are obviously fulfilled in our case. Therefore we can consider the Dingle temperature in Eq.\,(\ref{RD}) to be free of many-body renormalization but simultaneously replace the renormalized mass $\mu$, by the band mass $\mu^{\mathrm{b}}$. By doing that, we come to new values for the Dingle temperatures in our fits: $T_{\mathrm{D},\alpha} = 1.4 \pm 0.05$\,K and $T_{\mathrm{D},\beta} = 2.28\pm 0.005$\,K. One can see that the relative difference between them has reduced, however is still quite large and cannot be explained by the experimental error or uncertainties in the fitting procedure.

Thus, the assumption of a momentum-independent scattering time
$\tau$ seems to be inappropriate in our case.
As mentioned in Sect.\,\ref{Sect_mass}, the nesting property of the rhombus-like $\alpha$ orbit may cause enhanced scattering on this part of the Fermi surface. However, this should lead to a relative increase of $T_{\mathrm{D},\alpha} \propto 1/\tau_{\alpha}$, whereas our estimated value is considerably {\em lower} than $T_{\mathrm{D},\beta}$.

Another possibility is to consider a momentum-independent mean free
path $\ell$ instead of $\tau$ as a characteristic  parameter of
scattering for different states on the Fermi surface
\cite{comm_ell}. This may be a realistic scenario, for example,
if scattering is mainly determined by a 2D dislocation network \cite{shoe84}.
In Eq.\,(\ref{RD}) the scattering time can be replaced by the mean
free path with the help of the approximate relation
$\ell \approx \tau \overline{p}_{\mathrm{F}}/ m_{c}$, where
$\overline{p}_{\mathrm{F}} \simeq \sqrt{2e\hbar F}$ is the
relevant ``averaged'' Fermi momentum estimated from the SdH
frequency $F$.
Then, using the given above estimations of $T_{\mathrm{D},\alpha}$ and
$T_{\mathrm{D},\beta}$, we obtain $\ell_{\alpha} \simeq 97$\,nm and
$\ell_{\beta} \simeq  77$\,nm for the $\alpha$ and $\beta$ orbits, respectively.
These  two values are much closer to each other than
 $T_{\mathrm{D},\alpha}$ and $T_{\mathrm{D},\beta}$. This is
 because the relatively large $\tau_{\alpha}$ is partially compensated by the
strongly enhanced effective cyclotron mass $m_{c,\alpha}$, see
 Sect.\,\ref{Sect_mass}. Of course, these are only rough
estimates, taking into account the approximations made above.
However, one can consider this result as a hint to an important role
of dislocations in the damping of SdH oscillations in the
present material.

\section{Semiclassical magnetoresistance}
\label{Sect_MR}
\subsection{Field perpendicular to layers: effects of magnetic breakdown and field-induced dimensional crossover}
\label{Sect_Bperp}
\begin{figure}[tb]
    \centering
        \includegraphics[width=0.45\textwidth]{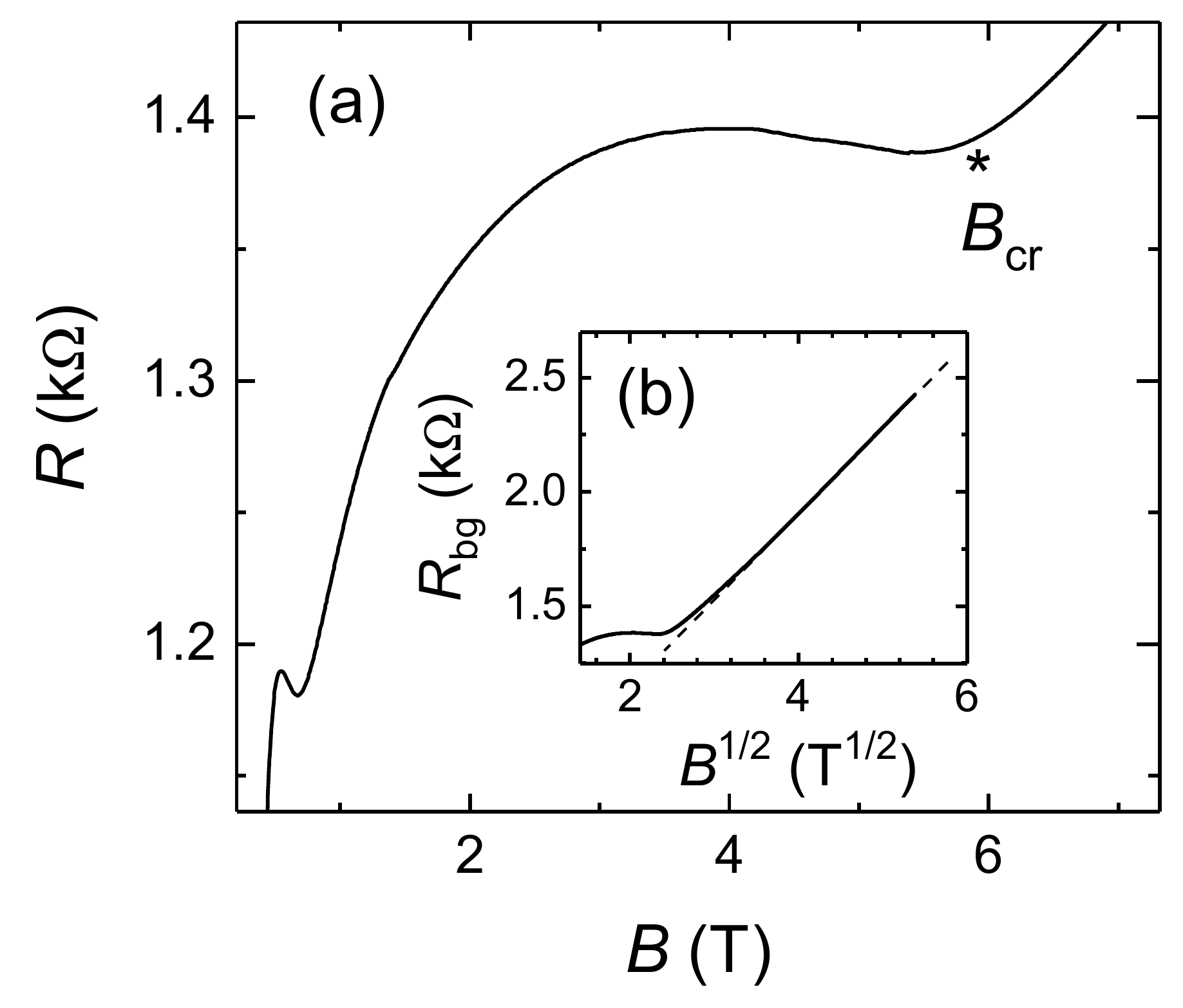}
    \caption{(a) Close-up of the field-dependent resistance data from Fig.\,\ref{R(B)} for
    fields $B< 7$\,T. The star marks the crossover between different magnetoresistance regimes,
    see text. (b) The nonoscillating resistance component plotted against square-root of
    magnetic field. Starting from $\sim 10$\,T the resistance acquires the
    $\sqrt{B}$-dependence. The dashed straight line is a guide to the eye.}
    \label{R_bg}
\end{figure}

Apart from the quantum oscillations,
the magnetoresistance of $\kappa$-(BETS)$_2$\-Mn\-[N(CN)$_2$]$_3$
exhibits several other features, which can be seen in detail
in Fig.\,\ref{R_bg}.
At zero field the material is superconducting. However,
the zero-resistance superconducting state is very rapidly suppressed
in this field geometry: the normal-state resistance is already
restored at $B \approx 0.5$\,T.  The resistive transition is
followed by a small sharp peak. This anomaly is better pronounced at liquid
$^{4}$He temperatures \cite{zver10} and has been observed on a
number of other layered organic superconductors
\cite{zuo96,su99,yama96}. Its origin is most likely associated with
a specific influence of superconducting fluctuations on
the interlayer conduction in a strongly anisotropic,
quasi-2D superconductor \cite{frie97,kart99,glat11}.

In the fully normal state the resistance begins to increase rapidly with
the field. However, already starting from $\sim 1.2$\,T it gradually
flattens out, shows a broad maximum around 4\,T and slightly, by $\sim 1\%$,
decreases as the field is further increased to $\approx 5.5$\,T.

One can qualitatively explain this behavior in terms of the MB effect.
Generally speaking, at low fields, $B \ll B_{\mathrm{MB}}$, there are two distinct types of
electron orbits on the Fermi surface: closed orbits on the $\alpha$ pockets
and the orbits on the open Fermi sheets. The contribution of the carriers
on the closed orbits to the interlayer conductivity only weakly depends
on magnetic field, whereas the contribution from the open orbits
decreases proportionally to $1/B^2$ \cite{lifs57,lifs58,abri88}.
Thus, in the absence of MB, the open orbits are ``freezing out'' and
magnetoresistance rapidly increases with field, asymptotically
approaching a value solely determined by the closed cyclotron
orbits. When the field becomes comparable to $B_{\mathrm{MB}}$,
tunneling of carriers through the MB gap gives rise to new closed
orbits and reduces the relative weight of the classical open orbits.
Finally, at $B \gg B_{\mathrm{MB}}$ almost all the carriers execute
the large closed $\beta$ orbit, equally contributing to the
interlayer conductivity. This leads to a significant increase of
interlayer conductivity, hence, decrease of resistivity in
comparison to what it would be without MB.

While the above description is only qualitative, one could try to
roughly estimate the MB field, ascribing it to the field at which the
magnetoresistance considerably curves down from its initial slope.
According to Fig.\,\ref{R_bg}(a), it happens in the interval between
$1.5$ and $3$\,T, which would imply a MB gap
$\Delta_{\mathrm{MB}} \simeq 1.3 - 2$\,meV.
This is much higher than $0.3$\,meV obtained from the analysis
of the quantum oscillations.
However, as pointed out in Sect.\,\ref{Sect_B}, the latter value is
also only a very rough estimation. Therefore, the obtained discrepancy
should not be considered as a severe contradiction between two methods
but rather as an illustration of subtlety of such kind of estimations.

At high enough fields, $B \gg B_{\mathrm{MB}}$, the large closed
orbit $\beta$ dominates. However, the magnetoresistance does not keep
saturating, as the standard magnetotransport theory
\cite{lifs57,lifs58,abri88} predicts, but rather grows notably
starting from $B_{\mathrm{cr}} \simeq 6$\,T all the way up to the highest field.
Above 10\,T it precisely follows a $\sqrt{B}$-dependence, as
illustrated in Fig.\,\ref{R_bg}(b). The
same field dependence has recently been reported for another highly
anisotropic layered conductor and explained in terms of a
field-induced dimensional crossover to a ``weakly coherent''
interlayer transport regime \cite{grig12}. The latter is defined as
coherent interlayer charge transfer under a strong magnetic field,
when the cyclotron frequency $\omega_c$ significantly exceeds
both the zero-field intralayer scattering rate $1/\tau_0$ and the interlayer
tunneling rate $\sim t_{\perp}/\hbar$. At such conditions the impact of scattering on point-like
defects on charge transport is enhanced
similarly to the case of a purely two-dimensional system
\cite{ando74a}. As a result, the scattering rate and hence interlayer resistance are
predicted \cite{grig11,*grig11a,grig13} to grow proportionally to $\sqrt{B}$.

Assuming that bending of the $R(B)$ dependence at $B_{\mathrm{cr}}$ is
entirely caused by the crossover to the weakly coherent regime, one
can evaluate the upper limit for the interlayer transfer integral:
$2t_{\perp} \leq \hbar\omega_{c,\mathrm{cr}}
\equiv \hbar eB_{\mathrm{cr}}/m_{c,\beta}
= 0.1$\,meV (here we used the value for the cyclotron mass
$m_{c,\beta} = 7.0 m_e$ obtained from the SdH data).
Estimating, further, the Fermi energy as
$\varepsilon_{\mathrm{F}} \sim \hbar e
F_{\beta}/m_c \approx 70$\,meV, we obtain the anisotropy ratio,
$\varepsilon_{\mathrm{F}}/2t_{\perp} \geq 700$.
The information on $t_{\perp}$ in known BETS salts is very scarce.
Comparing with better studied $\kappa$-type BEDT-TTF salts, similar
values have been reported for the most anisotropic compounds
$\kappa$-(BEDT-TTF)$_2$X with X = Cu(NCS)$_2$ \cite{sing02}
and I$_3$ \cite{wosn02}. The very low value of $t_{\perp}$ is
fully consistent with the absence of beats in the SdH oscillations
and thus further justifies the use of the 2D Lifshitz-Kosevich-Shoenberg
formula (\ref{Ai}) for the oscillation amplitude.

In Figs.\,\ref{R(B)} and \ref{R_bg}, the bending of magnetoresistance
at $B_{\mathrm{cr}}$
may look somewhat too sharp for what one would generally expect from
a gradual crossover.
However, a similar sharp bending has been
found on pressurized $\alpha$-(BEDT-TTF)$_2$KHg(SCN)$_4$ at
$\simeq 1$\,T \cite{grig12}.
Moreover, the recent numerical calculations of the interlayer
magnetoresistance in a quasi-two-dimensional metal \cite{grig14}
made in the self-consistent Born approximation have reproduced
a relatively sharp crossover from a nearly constant value to a
$\sqrt{B}$-dependence in a field when the Landau level separation becomes
larger than their width $2\Gamma_0 = \hbar/\tau_0$.
Note that in this case the crossover is
determined by scattering rather than by interlayer charge
transfer: the condition $\hbar\omega_{c} \gg 2t_{\perp}$ is supposed to be fulfilled
already at lower fields. If we adopt the same crossover criterion,
$\hbar\omega_{c,\mathrm{cr}} \simeq 2\Gamma_0$, for our case, we can estimate the
transport scattering time, $\tau_0 \simeq m_{c,\beta}/eB_{\mathrm{cr}}\approx 6.5$\,ps.
This value is much higher than the scattering time derived from the Dingle factor of
the SdH oscillations,
$\tau_{\mathrm{D}} = \hbar/2\pi k_{\mathrm{B}}T_{\mathrm{D}} \approx 0.6$\,ps.
Such a large difference is not uncommon for organic metals \cite{hill97,kart02a,grig12}
and is caused by different scattering mechanisms dominating in the charge
transport and quantum oscillation damping. In particular, as argued in Sect.\,\ref{Sect_B},
the Dingle factor in our material is mainly determined by scattering on dislocations, which
usually plays only a minor role in the charge transport, especially in the interlayer direction.

\subsection{Angle-dependent magnetoresistance and the Fermi surface geometry}
\label{AMR}
\begin{figure}[tb]
    \centering
        \includegraphics[width=0.48\textwidth]{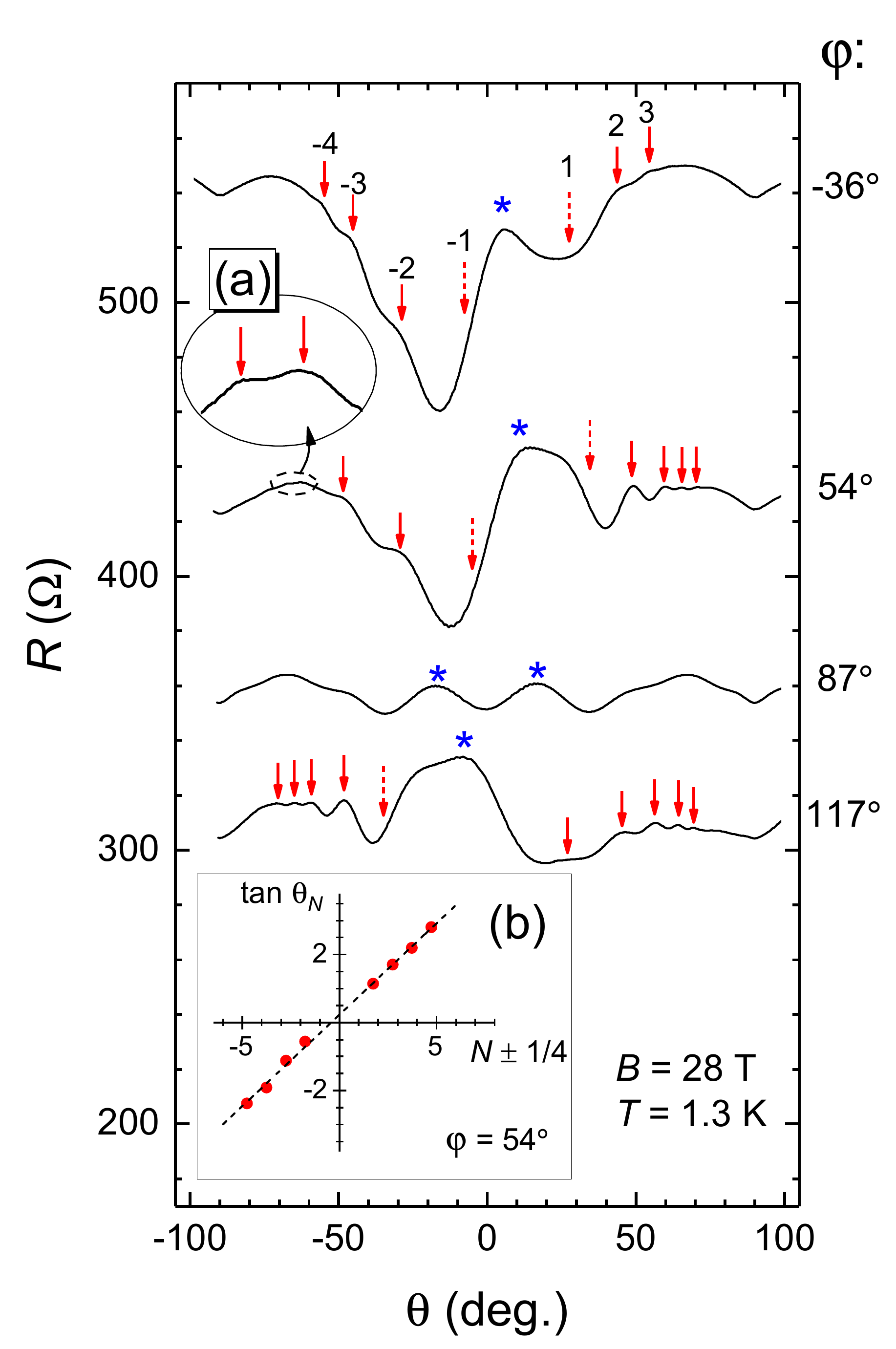}
    \caption{(Color online) Examples of the angular dependence of the resistance in a magnetic field
    of 28\,T at different azimuthal orientations $\varphi$ indicated on the right-hand side.
    The curves are vertically shifted for clarity.
    The arrows point to the positions of AMRO maxima, see text. The stars mark the -1st and +1st maxima
    corresponding to additional oscillations dominating near $\varphi \simeq 90^{\circ}$.
    Insets: (a) high-$\theta$ part of the $R(\theta)$ dependence at $\varphi= 54^{\circ}$ in an
    enlarged scale, showing high-order AMRO maxima; (b) AMRO positions plotted in the $\tan \theta$ vs. $N$ scale, for
    $\varphi= 54^{\circ}$. The  positive and negative indices $N$ are shifted, respectively, to the left
    and to the right by $1/4$, in order to enable a common linear fit according to Eq.\,(\ref{yamaji}).}
    \label{AMRO}
\end{figure}

The main panel of Fig.\,\ref{AMRO} shows examples of the resistance recorded at
rotating the sample in a constant magnetic field, $B=28$\,T. The resistance was measured
as a function of polar angle $\theta$ at different fixed azimuthal angles $\varphi$,
see Sect.\,\ref{exper}. The measurements were done in the range
$-36^{\circ} \leq \varphi \leq 144^{\circ}$.
For most of the azimuthal orientations clear oscillations periodic in $\tan \theta$ have been
detected.
In Fig.\,\ref{AMRO} the relevant local maxima are marked by arrows.
These angle-dependent magnetoresistance oscillations (AMRO), also known as Yamaji oscillations,
originate from periodic geometric resonances of the interlayer charge transport in a quasi-2D metal in a
tilted magnetic field \cite{yama89,yagi90a,pesc91,kuri92,yako06,grig10}. They are frequently
observed in organic metals and utilized for exploring the Fermi surface geometry \cite{kart04}.

To determine the shape of the Fermi-surface inplane cross section, we follow the procedure
proposed in Ref.\,\citenum{kart92a} for the general case of a low-symmetry layered system.
First, for each $\theta$-sweep the AMRO period is evaluated from the
linear fit of the $N$-th local maximum positions plotted in the $\tan\theta$-scale against $N$,
according to the condition:
\begin{equation}
\left| \tan\theta_N \right| = \Delta_0\left(\left|N + \gamma\right|-1/4\right), \qquad N=\pm 1,\pm 2, ...,
\label{yamaji}
\end{equation}
where the offset $\gamma$ ($-1< \gamma <1$) is determined by the inplane projection of the interlayer
hopping vector \cite{kart92a}.
An example of such fitting for $\varphi = 54^{\circ}$ is shown in Fig.\,\ref{AMRO}(b).
In the figure, the positive and negative indices are shifted by $-1/4$ and $+1/4$, respectively, in order
to include data for both positive and negative angles in a common fit.
\begin{figure}[tb]
    \centering
        \includegraphics[width=0.45\textwidth]{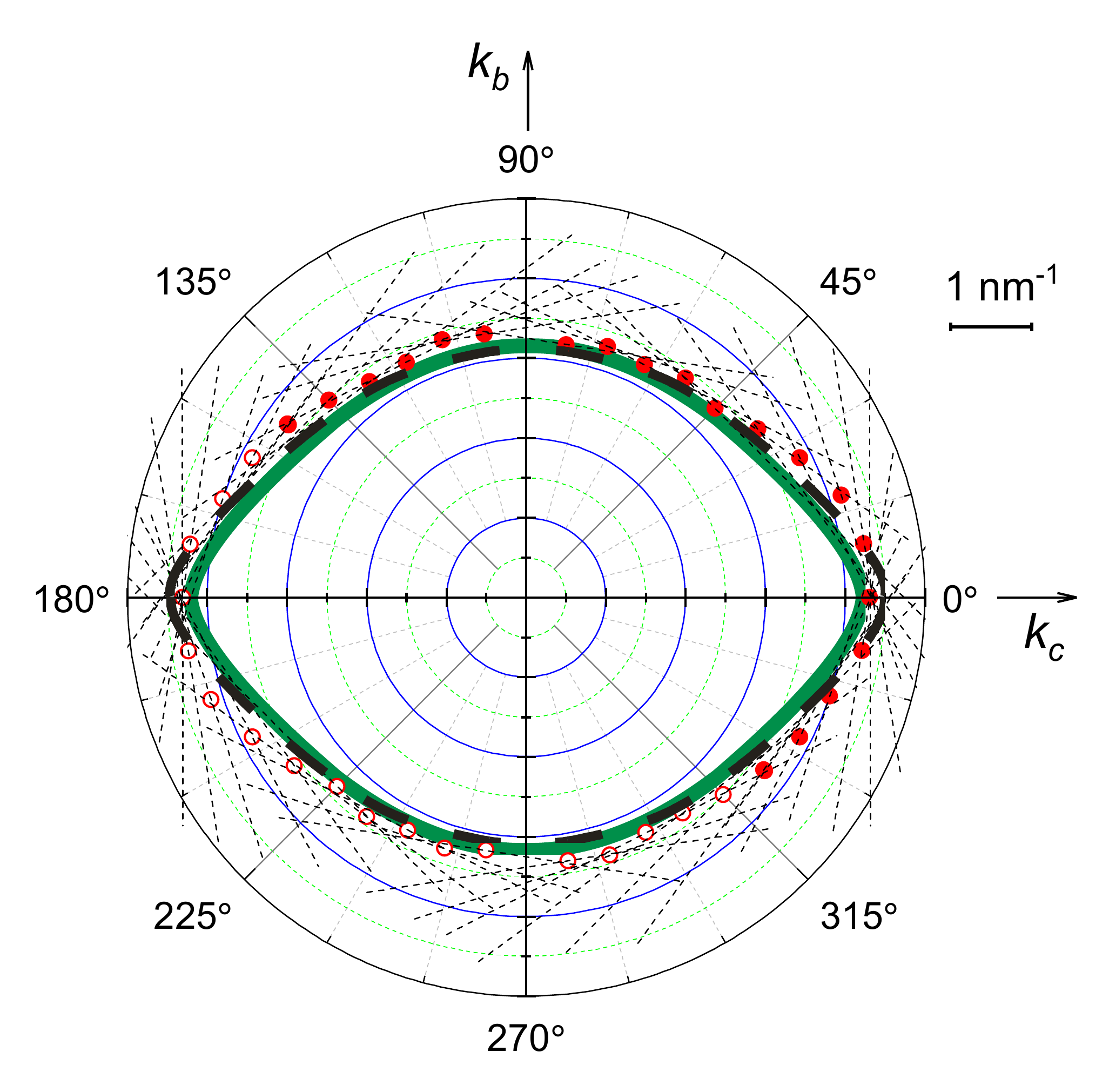}
    \caption{(Color online) Inplane cross section of the Fermi surface determined from the AMRO data
    as described in the text. The red symbols are the values of $k_B^{\mathrm{max}}(\varphi)$ in polar
    coordinates. Filled symbols are the data obtained directly from the $R(\theta)$ curves at the
    corresponding angles $\varphi$; open symbols are the same data shifted by $180^{\circ}$. For a
    comparison with the theoretical predictions the Fermi surface from Fig.\,\ref{FS} is shown by the thick
    dashed line.}
    \label{FS_AMRO}
\end{figure}

The period $\Delta_0$ is given by the ratio of the reciprocal lattice period $K_z$ in the interlayer
direction and the maximum inplane Fermi wave vector projection $k_B^{\mathrm{max}}$ on the field
rotation plane: $\Delta_0(\varphi) = 1/2 \left[K_z/k_B^{\mathrm{max}}(\varphi)\right]$. Thus, using the
experimentally determined AMRO periods and substituting $K_z=3.24$\,nm$^{-1}$ taken from
the low-temperature crystallographic data \cite{zver10}, we evaluate $k_B^{\mathrm{max}}$
for different azimuthal orientations $\varphi$. The result, in polar coordinates,
is shown in Fig.\,\ref{FS_AMRO}. Here, the filled circles represent the data obtained directly
from the experiment, while
the open circles are the same data translated
by $180^{\circ}$, taking into account the inversion symmetry of the Fermi surface. Further, through
each point $k_B^{\mathrm{max}}(\varphi)$ a straight line perpendicular to the direction $\varphi$ is
drawn (thin dashed lines in Fig.\,\ref{FS_AMRO}) and
the inplane Fermi surface (thick green line) is constructed as a contour inscribed in the whole
set of these straight lines.

The size of the obtained Fermi surface is close to that of the first Brillouin zone, in agreement
with the large Fermi surface predicted by the band structure calculations and with the frequency
of the fast quantum oscillations presented above. No AMRO associated with the small $\alpha$ pockets of
the Fermi surface have been found. This is obviously a consequence of the strong MB regime which
governs the magnetoresistance behavior at $B=28$\,T even at relatively high tilt angles, at least
up to $\simeq 70^{\circ}$.

As to the shape of the Fermi surface, it is also quite similar to the theoretical
one, which is indicated in Fig.\,\ref{FS_AMRO} by the thick dashed line. It also has relatively flat
segments inclined by $\simeq \pm 40^{\circ}$ with respect to the $k_c$-axis and a sharp ``nose'' along
$k_c$. The dimension along $k_c$ appears to be slightly smaller and along $k_b$ slightly larger than
calculated. However, the difference does not exceed the experimental error bar. Therefore we can speak
about very good quantitative agreement between the theoretical predictions and the experimental results.

Surprisingly, at azimuthal orientations in a narrow interval of $\simeq \pm 5^{\circ}$ around $k_b$
direction the conventional AMRO vanish. In this interval the $R(\theta)$ dependence
is governed by other, nonperiodic oscillations, see the $\varphi = 87^{\circ}$ curve in
Fig.\,\ref{AMRO}. The new features seem to compete with the AMRO.
Outside the mentioned $\varphi$ range they only persist at low tilt angles, $|\theta| < 30^{\circ}$.
At the same time the AMRO maxima, which are expected at the positions pointed by
dotted arrows in Fig.\,\ref{AMRO}, are completely suppressed. By contrast, at higher $\theta$
the conventional AMRO are restored, whereas the new features disappear. Thus, there seems to be no angular
range where both kinds of oscillations coexist. It should be noted that at the angles, at
which switching between the two kinds happens, no change in the cyclotron orbit topology is expected.

Similarily to the AMRO and other geometrical effects of the Fermi surface in a quasi-2D metal \cite{kart04},
the new features appear to keep their angular positions at changing the field strength.
\begin{figure}[tb]
    \centering
        \includegraphics[width=0.45\textwidth]{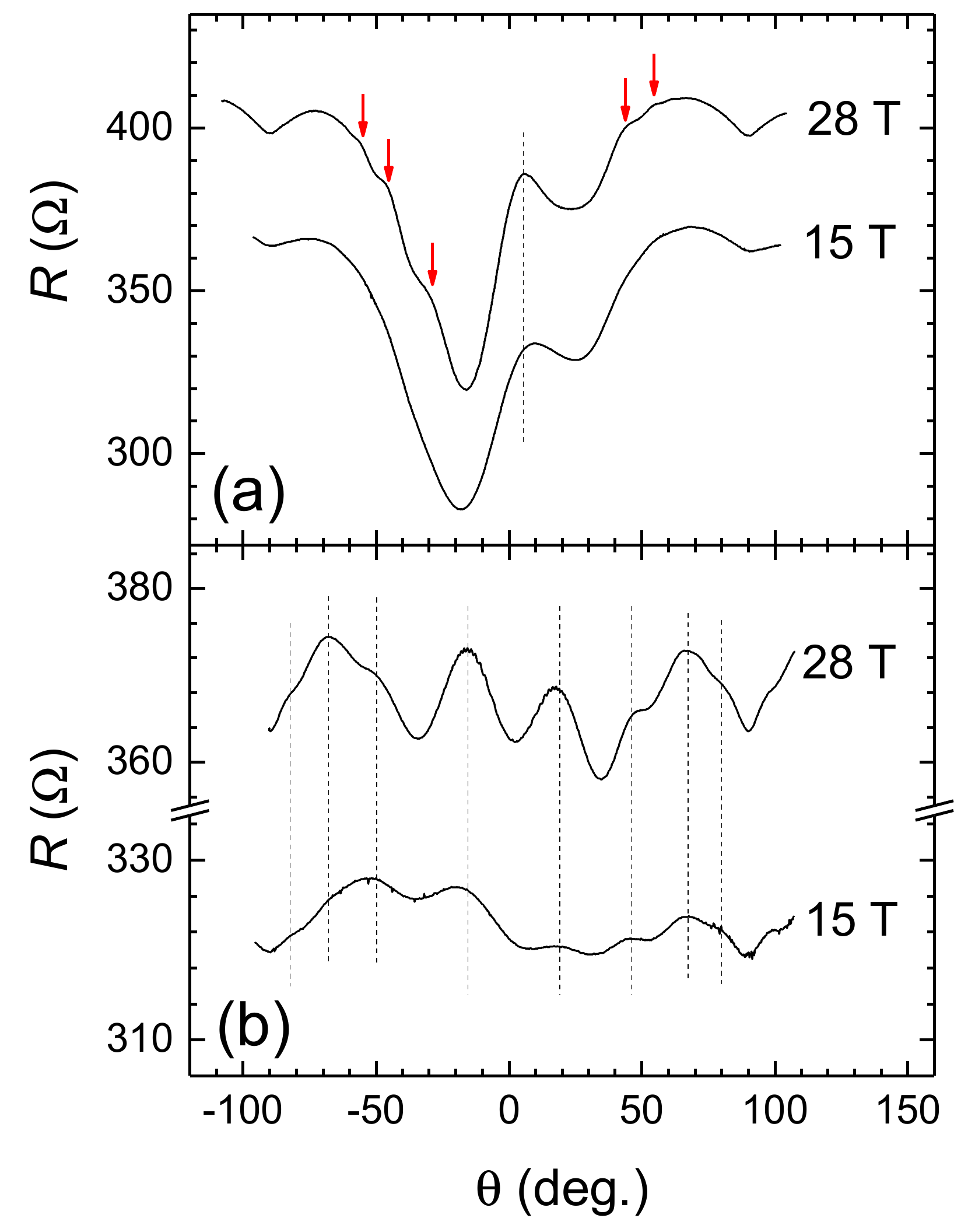}
    \caption{$\theta$-sweeps at $B= 15$ and 28\,T, at: (a) $\varphi = -36^{\circ}$ and (b) $\varphi = 93^{\circ}$.
    Arrows point to the AMRO positions. Dashed vertical lines illustrate the independence of the positions of
    the additional, ``non-AMRO'' features on the field strength.}
    \label{AMR_B}
\end{figure}
This is illustrated in Fig.\,\ref{AMR_B} where the $R(\theta)$ dependence is shown for two azimuthal
orientations at $B=28$ and 15\,T. On the other hand, the influence of the field strength on the amplitude
of the features is relatively weak: Fig.\,\ref{AMR_B}(a) shows that at decreasing the field to 15\,T
the usual AMRO practically vanish, whereas the peak around $\theta = 5.5^{\circ}$ only becomes slightly
lower and more smeared.

All in all, the new features are unlikely a pure effect of the Fermi surface geometry.
One may look for their origin in coupling between charge and spin degrees of freedom. Indeed, the
magnetoresistance, especially in the interlayer direction, may be sensitive to the magnetic state
of Mn$^{2+}$ ions in the anion layer. The ambient-pressure magnetic experiments \cite{vyas11,vyas12,vyas17}
have revealed a dramatic slowing of the spin dynamics in the manganese subsystem at low temperatures
and considerable interactions with antiferromagnetically ordering $\pi$-electron spins in the
Mott-insulating state. However, at present we do not have enough data to establish a direct link between
the magnetoresistance behavior and magnetic properties. A further study, for example, combined
angle-resolved resistance and magnetic measurements in the metallic state should be helpful for
clarifying the situation.

\begin{figure}[tb]
    \centering
        \includegraphics[width=0.45\textwidth]{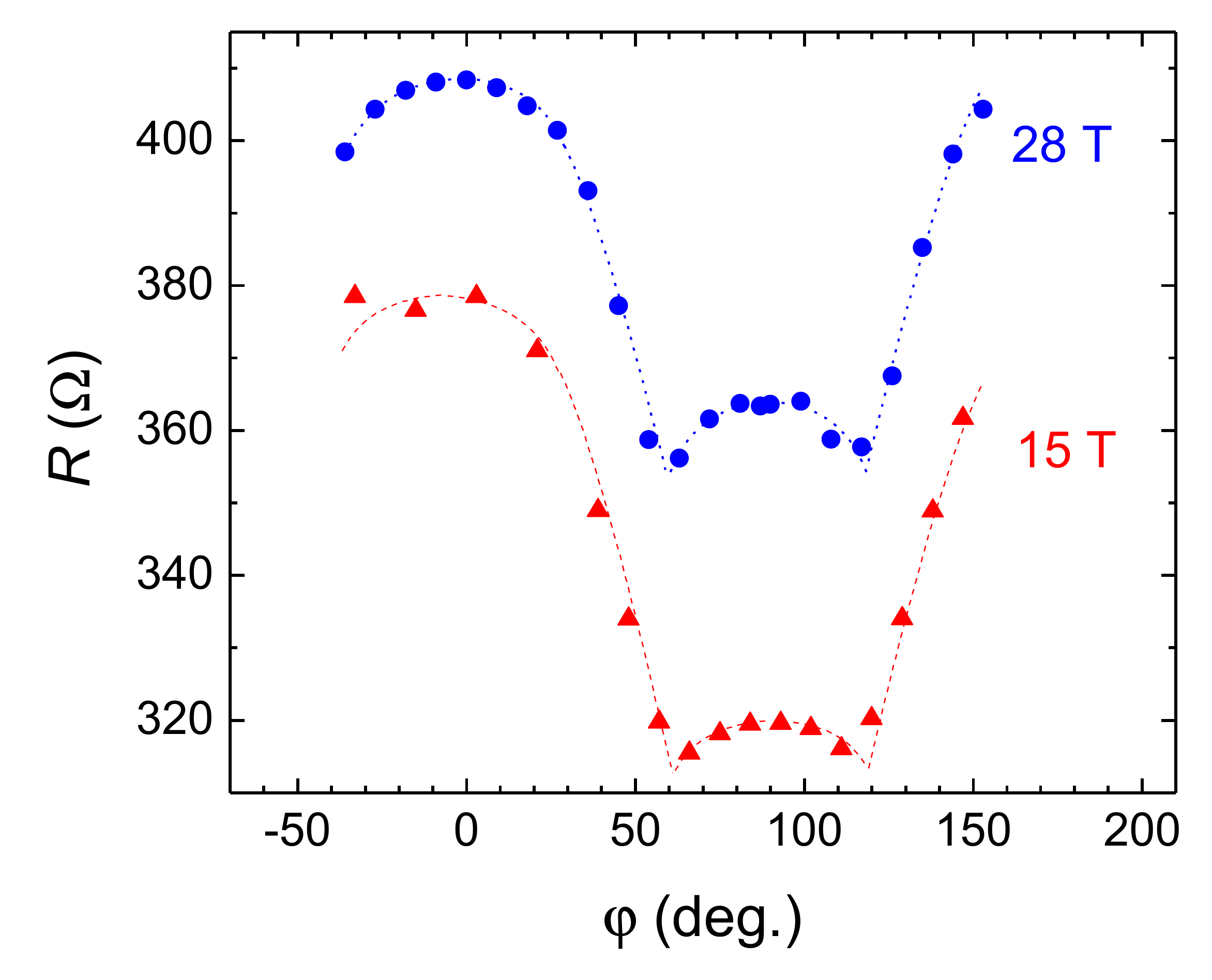}
    \caption{(Color online) Angular dependence of the resistance in a field parallel to the conducting layers,
    at $B= 15$ and 28\,T. Dotted lines are guides to the eye.}
    \label{inplane}
\end{figure}

Finally, we discuss the dependence of magnetoresistance on the inplane field orientation.
Figure\,\ref{inplane} shows the resistance values corresponding to $\theta = 90^{\circ}$, taken
from the $R(\theta)$ curves recorded at different $\varphi$.
The notable variation of the resistance, as the field is turned in the layer plane, is
generally associated with coherent interlayer charge transport in a layered system with an
anisotropic inplane  cross section of the Fermi surface \cite{kart05b,kart06}

The shape of the $R(\varphi)$ dependence resembles that observed on materials with open Fermi sheets
such as Bechgaard salts (TMTSF)$_2$X \cite{osad96,lee98b,kang09} or (DMET)$_2$X \cite{yosh95,yosh97}.
The magnetoresistance is at a maximum when the field is directed along the plane of the sheets, in our
case $\varphi = 0^{\circ}$ (i.e. $\mathbf{B} \| \mathrm{c}$), and decreases towards $\varphi = 90^{\circ}$,
exhibiting two minima at $\varphi \approx 90^{\circ} \pm 30^{\circ}$.
As shown in Fig.\,\ref{inplane}, a change of the field strength by almost a factor of two only affects the
absolute value of magnetoresistance; the shape of the $\varphi$-dependence remains largely unchanged. Most
importantly, the angles of the resistance minima stay the same, suggesting that they are associated with the
Fermi surface geometry. At first glance, they can be ascribed to the ``third angular effect''at the field
directions perpendicular to inflection points on the Fermi surface \cite{osad96,lee98b}.
However, our Fermi surface contains, besides the open sheets,
a cylindrical $\alpha$-pocket. Moreover, since the gaps between the sheets and the pocket are very small,
it is the large closed Fermi surface
cross section delineated in Fig.\,\ref{FS_AMRO}, which basically determines the magnetoresistance behavior.
Assuming a momentum-independent scattering rate, one would expect minima of $R(\varphi)$ at a field perpendicular
to the parts of the Fermi surface with the smallest curvature \cite{kart04}, that is,
perpendicular to the flat segments of the $\alpha$-pocket. According to Figs.\,\ref{FS} and
\ref{FS_AMRO}, these directions are tilted by $\approx 40^{\circ}$ from the $b$-axis, i.e. $10^{\circ}$
away from the positions of the detected resistance minima. This discrepancy definitely exceeds the experimental
error bar and the uncertainty in the Fermi surface shape.

A shift of the resistance minimum positions from the directions given by the Fermi surface geometry may
occur if the carrier mobility significantly depends on momentum
For example, Sugiwara et al. \cite{sugi13a,sugi13b}
used the apparent mismatch between the $R(\varphi)$ dependence and the Fermi surface shape to evaluate the
$\mathbf{k}$-dependence of scattering rate in some BETS and BEDT-TTF salts.
In our case, the shift of the resistance minima towards the direction of the $b$-axis can be explained by
a suppressed mobility on the $\alpha$ Fermi pocket. This seems to be a likely scenario, keeping in mind the
particularly strong enhancement of the effective mass $m_{c,\alpha}$ revealed in the SdH experiment.

As mentioned above, the significant $\varphi$-dependence of magnetoresistance associated with the Fermi surface
geometry is an evidence of a coherent interlayer charge transport. On the other hand,
a broad dip observed in all $\theta$-sweeps around $\theta = \pm 90^{\circ}$, see, e.g., Fig.\,\ref{AMRO},
reveals the presence of an incoherent conduction channel \cite{kart06,kart09b}.
Thus, the total interlayer conduction includes both
the coherent and incoherent channels. However, the small amplitude of the dip ($< 10\%$ of the total
magnetoresistance) indicates that only a minor fraction of the total conductivity is incoherent.

Despite the dominant contribution of the coherent conduction channel, we were unable to detect
a sharp peak in the angle-dependent magnetoresistance around $\theta = 90^{\circ}$ \cite{pesc99a,hana98},
which is often considered as a fingerprint of the coherent interlayer transport regime.
The reason for that is
most likely the very high anisotropy. Indeed, besides the effective field-strength parameter $\omega_c\tau$,
the magnitude of such ``coherence peak'' depends on the anisotropy ratio \cite{hana98}. The peak is particularly strong
in clean quasi-2D metals with a moderately high anisotropy, $\varepsilon_{\mathrm{F}}/2t_{\perp} \sim 100-200$
\cite{hana98,kart92a,ohmi99,kiku16} but diminishes in more anisotropic materials \cite{wosn02,kart06,sing02}.
In a clean sample of $\kappa$-(BEDT-TTF)$_2$Cu(NCS)$_2$, showing an anisotropy similar to our compound, the
peak height was found to be only few percent of the total resistance in fields $42-45$\,T \cite{sing02,godd04}.
In our experiment the highest field was 1.5 times lower and the Dingle temperature about 4 times higher
($T_{\mathrm{D}}\simeq 2$\,K against 0.5\,K in Ref.\,\citenum{godd04}), which explains the absence
of the peak in the present data.

\section{Summary}
\label{summary}
We have studied the Fermi surface properties of $\kappa$-(BETS)$_2$Mn[N(CN)$_2$]$_3$
both theoretically and experimentally.
The experiment was done under a moderate pressure,
in order to stabilize the metallic ground state. The large cylindrical Fermi
surface predicted by the calculations is found to be split into a pair of open sheets extended
in the $k_ak_c$-plane and a cylinder with a rhombus-like inplane cross-section
area of  $\approx 27\%$ of the first Brillouin zone. The sheets and the cylinder are separated from
each other by small, of the order of 1\,meV, gaps on the Brillouin zone boundary caused by a disorder of the
dicyanamide groups in the anion layers and, possibly, by a weak spin-orbit interaction.
The predicted size and shape of the Fermi surface are confirmed by the SdH oscillations as well as
by the classical AMRO.
In particular, the SdH spectrum shows two fundamental frequencies
corresponding, respectively, to the classical cyclotron orbit $\alpha$ on the small Fermi cylinder
and to the large $\beta$ orbit caused by magnetic breakdown through the gaps.
While the topology of the Fermi surface is typical of the $\kappa$-type salts of BEDT-TTF and
BETS, there are a few interesting features specific to the present compound.

The effective cyclotron masses determined from the $T$-dependence of the SdH oscillations show a
strong enhancement which can only partially be attributed to the peak in the calculated one-particle DOS
near the Fermi level. The main reason for the enhancement is the renormalization effect of electron
correlations in the vicinity of the Mott-insulating transition.
The effect clearly exhibits a
momentum dependence, being especially strong on the $\alpha$-pocket of the Fermi surface.
A likely reason for the momentum-selective enhancement of correlations is a partial flattening of
the conducting band, associated with the $\alpha$-pocket, which places this part of the system
more close to the metal-insulator transition.
Additionally, the nesting property of this pocket could contribute to the instability of the metallic
state. The heavier effective mass leads to a lower mobility of the carriers on the $\alpha$-pocket. Indeed,
a close examination of the $\varphi$-dependence of the classical magnetoresistance in conjunction with the
shape of the inplane Fermi surface indicates a suppressed contribution of these carriers to the interlayer
conductance.
Further experiments at different pressures are required in order to understand how the
insulating instability develops on different parts of the Fermi surface upon approaching the metal-insulator
transition.

The field dependence of the SdH amplitude reveals a considerable difference between the Dingle temperatures
corresponding to the $\alpha$ and $\beta$ orbits. The apparent contradiction can be solved by suggesting a
constant mean free path instead of scattering time to be the relevant parameter in the Dingle factor. This
is a realistic scenario if the Landau level broadening responsible for damping of the oscillations is
mainly determined by scattering on a 2D dislocation network.
By contrast to the quantum oscillations, the classical magnetoresistance is largely insensitive to dislocations.
As a result, the scattering time estimated from the crossover in the magnetoresistance field dependence is
considerably longer than that inferred from the quantum oscillations.

The crossover field also sets the
upper limit for the interlayer transfer energy, $2t_{\perp} \leq 0.1$\,meV, which is $\simeq 700$ times lower
than the Fermi energy. Despite the weak coupling between the layers, the interlayer charge transport is dominated
by the coherent conduction channel. This is indicated both by prominent AMRO and by the considerable
$\varphi$-dependence of the magnetoresistance in a field parallel to layers.

Besides the well-known AMRO effect, an additional series of pronounced nonmonotonic features has been found in the
$\theta$-dependent magnetoresistance. These features are unlikely a pure effect of the Fermi surface geometry.
Keeping in mind the proximity to the insulating state with nontrivial magnetic properties,
they might be a result of charge-spin coupling in the presence of a magnetic instability.
Ideally, combined resistive and magnetic studies under pressure should clarify this point. This is, however,
a difficult experimental task due to the very small size of the samples. On the other hand, one could gain
additional information from a magnetoresistance study of the mixed salt
$\kappa$-(BETS)$_2$Co$_{0.13}$Mn$_{0.87}$[N(CN)$_2$]$_3$. This sister compound displays very similar phase diagram
and zero-field resistive properties but a considerably different magnetic anisotropy as compared to the present
salt \cite{kush17}. Confronting the magnetoresistance behaviors of the two salts may be helpful for understanding
the origin of the new oscillations.

\section{Acknowledgements}

We are grateful to N.D. Kushch for providing the high-quality crystals for
our studies and to P.D. Grigoriev for numerous useful discussions.
The work was supported by the German Research Foundation (DFG) via the 
grant KA 1652/4-1. 
The high-field measurements were done under support of the LNCMI-CNRS, 
member of the European Magnetic Field Laboratory (EMFL). 
V.N.Z. acknowledges the support from RFBR Grant 18-02-00280. 
Work in Spain was supported by the Spanish Ministerio de Economía y 
Competitividad (Grants FIS2015-64886-C5-4-P and CTQ2015-64579-C3-3-P) 
and Generalitat de Catalunya (2017SGR1506,  2017SGR1289  and XRQTC). 
E.C. acknowledges support from the Severo Ochoa Centers of Excellence 
Program under Grant SEV-2015-0496. P.A. acknowledges support from 
the Maria de Maeztu Units of Excellence Program under Grant MDM-2017-0767."


\begin{thebibliography}{105}
\bibitem [{com({\natexlab{a}})}]{comment_pressure}%
  \BibitemOpen
  \href@noop {} {} \bibinfo {note} {We estimate the
  renormalization factor $r$ by comparing the effective cyclotron masses
  determined at pressures $p = 1.4$ and $4.1$\,kbar with the theoretical values
  obtained from the ambient-pressure band structure. This is a reasonable
  approximation, since the expected pressure-induced relative changes in the
  band masses do not exceed a few percent within the given pressure
  range}\BibitemShut {NoStop}%
\bibitem [{com({\natexlab{b}})}]{comm_ell}%
  \BibitemOpen
  \href@noop {} {} \bibinfo {note} {Of course, we consider
  the mean free path in the conducting layers, since it is the inplane electron
  dynamics, which determines the behavior of magnetic quantum oscillations in
  our quasi-2D metal.}\BibitemShut {Stop}%
\end{thebibliography}
%

\end{document}